\documentclass[final,5p,times, twocolumn, sort&compress]{elsarticle}
\pdfoutput=1

\usepackage{amssymb}
\usepackage{amsthm}
\usepackage{amsmath}
\usepackage{mathtools}

\renewcommand{\phi}{\varphi}
\renewcommand{\epsilon}{\varepsilon}
\usepackage{chemformula}

\usepackage{siunitx}
\usepackage[font=small,labelfont=bf,labelsep=endash,format=plain]{caption}
\usepackage{subcaption}

\usepackage{tikz}					
\usetikzlibrary{matrix, shapes.geometric, arrows, positioning, external, decorations.fractals,%
	decorations.shapes,%
	decorations.text,%
	decorations.pathmorphing,%
	decorations.pathreplacing,%
	decorations.footprints,%
	decorations.markings}
\tikzstyle{io} = [trapezium, trapezium left angle=70, trapezium right angle=110, minimum width=3cm, minimum height=1cm, text centered, draw=black, fill=blue!30]
\tikzstyle{process} = [rectangle, minimum width=3cm, minimum height=1cm, text centered, text width=6.5cm, draw=black, rounded corners=0.2cm]
\tikzstyle{process2} = [rectangle, minimum width=3cm, minimum height=1cm, text centered, text width=3cm, draw=black, rounded corners=0.2cm]
\tikzstyle{decision} = [diamond, minimum width=3cm, minimum height=1cm, text centered, draw=black, fill=green!30]
\tikzstyle{arrow} = [thick,->,>=stealth]
\usepackage{varwidth}

\usepackage{pgf}

\usepackage{doi}					

\usepackage{xspace}

\usepackage[]{url} 					
\usepackage[]{hyperref}
\hypersetup{
	colorlinks,
}
\usepackage[nameinlink,capitalize]{cleveref}

\usepackage{float}

\usepackage{multirow}
\usepackage{booktabs}
\usepackage{tabularx,ragged2e}
\newcolumntype{C}{>{\Centering}X}    
\usepackage{array}
\usepackage{longtable}
\usepackage{rotating}

\usepackage{csquotes}
\usepackage{xpatch}
\usepackage{lmodern}
\usepackage[obeyFinal]{todonotes}
\usepackage{adjustbox}

\usepackage{tablefootnote}

\newcommand{\D}{\, \mathrm{d}}

\def\eg{e.g.,\xspace} 
\def\ie{i.e.,\xspace} 
\def\cf{cf.\xspace} 
\def\tca{t_{ca}\xspace}

\newcommand{\flp}{\emph{Flying Laptop}\xspace}

\renewcommand{\v}[1]{\boldsymbol{#1}}
\newcommand{\m}[1]{\boldsymbol{#1}}

\newcommand{\mat}[1]{\ensuremath{\begin{bmatrix} #1 \end{bmatrix}}}
\newcommand{\transp}[1]{\ensuremath{#1^{\text{T}}}}

\newcommand{\code}[1]{\texttt{#1}}

\journal{Acta Astronautica}

\begin{document}

\begin{frontmatter}


\title{An analysis tool for collision avoidance manoeuvres using aerodynamic drag}


\author[inst1]{F. Turco}

\affiliation[inst1]{organization={Institute of Space Systems, University of Stuttgart},
            addressline={Pfaffenwaldring 29}, 
            city={Stuttgart},
            postcode={70569},
            country={Germany}}

\author[inst1]{C. Traub}
\author[inst1]{S. Gaißer}
\author[inst1]{J. Burgdorf}
\author[inst1]{S. Klinkner}
\author[inst1]{S. Fasoulas}

\begin{abstract}
Aerodynamic collision avoidance manoeuvres provide an opportunity for satellites in Low Earth Orbits to reduce the risk during close encounters. 
With rising numbers of satellites and objects in orbit, satellites experience close encounters more frequently. 
Especially those satellites without thrusting capabilities face the problem of not being able to perform impulsive evasive manoeuvres. For satellites in Low Earth Orbits, though, perturbing forces due to aerodynamic drag may be used to influence their trajectories, thus offering a possibility to avoid collisions. 
This work introduces a tool for the analysis of aerodynamic collision avoidance manoeuvres. Current space-weather data are employed to estimate the density the satellite encounters. Achievable in-track separation distances following a variation of the ballistic coefficient through a change in attitude are then derived by evaluating an analytical equation from literature.
Considering additional constraints for the attitude, e.g., charging phases, and uncertainties in the used parameters, the influence of a manoeuvre on the conjunction geometry and the collision probability is examined.
The university satellite \flp of the University of Stuttgart is used as an exemplary satellite for analysis, which show the general effectiveness of evasive manoeuvres employing aerodynamic drag. First manoeuvring strategies can be deducted and the influence of parameter uncertainties is assessed. 

\end{abstract}




\end{frontmatter}


\section{Introduction}
The regime of Low Earth Orbits (LEO), \ie altitudes below \SI{1500}{\kilo\meter} \cite{Montenbruck.2001}, offers great possibilities for Earth observation, thermospheric investigations, and the global connectivity market. This draws the attention of various actors and leads to a sharp rise in satellite numbers.
Growing numbers of satellites (and debris) do not only pose a threat to functional satellites. Every collision produces a huge amount of debris objects, which in turn threaten other satellites. Analyses have shown that this could end in an avalanche-like process, referred to as the Kessler syndrome, rendering the LEO regime useless for future generations \cite{Kessler.1980,Kessler.1991,Kessler.2000}. This highlights the need to minimize collision risk. Besides active debris removal, an obvious option for functional satellites is the implementation of collision avoidance manoeuvres (CAMs) in case of a predicted close encounter with another object. 
Typically, such manoeuvres are performed with impulsive thrusters to deflect the satellite trajectory. Satellites without thrusting capabilities, on the other hand, need other strategies to evade potential collisions. In LEO, aerodynamic drag due to the remaining atmospheric particles represents significant natural perturbing forces. There has been research on how to use them to control and manoeuvre asymmetrically-shaped satellites, \eg for satellite formation flight \cite{Traub.2020}. 
Although achievable forces are magnitudes smaller than what is possible with chemical thrusters, satellite orbits can be measurably altered given enough time, which allows for the implementation of CAMs. Satellites without thrusters can, therefore, greatly benefit from CAMs using aerodynamic drag.\par

One such satellite is the \flp of the Institute of Space Systems, University of Stuttgart. It orbits the Earth in a nearly circular polar orbit at an altitude of $~\SI{600}{\kilo\meter}$.
From its launch into LEO in July 2017 until September 2022, the \flp{}'s operators received over 5000 warnings for over 150 close encounters from the Joint Space Operations Center (JSpOC), with the lowest predicted miss distance reported as \SI{29}{\meter}. Since the \flp{} does not possess any thrusters, it has no possibility to perform impulsive collision avoidance manoeuvres. Thus, so far the operators had to remain without action upon reception of a collision warning. As previously pointed out, though, the orbital altitude and the asymmetric shape of the satellite, in principle, allow control via aerodynamic drag. The \flp will be used as an example satellite in this work.

At first, fundamental concepts are covered. The developed analysis tool is then introduced in \cref{ch:methodology}. Lastly, results of exemplary analyses for the \flp are presented and discussed in \cref{sec:results_discussion}. The research presented here is based on the corresponding author's master thesis \cite{Turco.2022}.
\section{Fundamentals}

\subsection{Satellite aerodynamics}
\label{sec:satellite_aerodynamics}
Satellites in LEO are subject to a measurable perturbing force due to the remaining atmospheric particles, which impinge on the satellite surface. The aerodynamic forces acting on a satellite can be separated into lift and drag. Drag is defined as the component acting anti-parallel to the satellite velocity relative to the local atmosphere, whereas lift acts in perpendicular direction to that. The specific drag force $ \v{f}_D $ depends on the satellite's dimensionless aerodynamic drag coefficient $ C_D $, the reference area $ A_{ref} $, the satellite mass $ m $, the atmospheric density $ \rho $ and the satellite's velocity relative to the local atmosphere $ \v{v}_{rel} $ \cite{Vallado}:
\begin{equation}
	\v{f}_D = -\frac{1}{2} \rho \frac{C_D A_{ref}}{m} v_{rel}^2 \frac{\v{v}_{rel}}{v_{rel}}
	\label{eq:drag}
\end{equation}
Neglecting atmospheric winds, the relative velocity equals the satellite's orbital velocity.

Lift coefficients are usually considerably smaller than the respective drag coefficients. Although aerodynamic lift can be used for satellite control under the right conditions \cite{Traub.2020, Traub.2022, Omar.2013}, perturbing lift forces are not further pursued here.\par
The ballistic coefficient $ \beta $ is a dimensional coefficient measuring to which extent a satellite is affected by aerodynamic drag. It is defined as \cite{Vallado}
\begin{equation}
	\beta = \frac{m}{C_D A_{ref}}.
\end{equation}
Here, the inverse ballistic coefficient $\beta^*$ is used, since it shows a direct proportionality to the drag coefficient and the reference area:
\begin{equation}
	\beta^* = \frac{C_D A_{ref}}{m}
	\label{eq:ballistic_coeff}
\end{equation}
Satellites with higher inverse ballistic coefficients experience higher drag forces than such with low inverse ballistic coefficients.\par
In \cref{eq:drag}, atmospheric density and satellite velocity are not controllable and the satellite mass is constant for thrusterless satellites. If a satellite is asymmetrically-shaped control authority lies in the variation of the ballistic coefficient. By changing its attitude and thereby changing the effective cross-section and the drag coefficient, the experienced drag force can be regulated.\par

The atmospheric density may be obtained by employing one of various atmospheric models of differing complexity. The NRLSMSISE-00 model \cite{Picone.2002} considers time, location and solar as well as geomagnetic activity via the solar radio flux $ F_{10.7} $ and the 3-hourly planetary equivalent amplitude values $ a_p $ to deliver estimations for the atmospheric constitution, temperature and total mass density \cite{Vallado,Doornbos.2011}. The JB-2008 model \cite{Bowman.2008} utilizes further proxies for solar activity. Besides the observed solar radio flux at \SI{10.7}{\centi\meter} ($ F_{10.7} $), proxies have to be provided for \SIrange{26}{34}{\nano\meter} solar extreme ultraviolet (EUV) emission ($ S_{10} $), solar middle ultraviolet (MUV) radiation at \SI{280}{\nano\meter} ($ M_{10} $) and X-ray emission of the sun ($ Y_{10} $). The disturbance storm time ($ Dst $) is used as an index for geomagnetic activity.\par

The \citet{ISO1422} has defined different levels of solar and geomagnetic activity by specifying values for the various parameters, which can be useful to compare different scenarios of solar and geomagnetic activity. The respective indices can be found in \cref{tab:activity_levels}.

\subsection{Analytic estimation of achievable separation distance}
\label{sec:sep_distance}
\citet{Omar.2020} derived an analytic expression for the evolution of the difference in mean anomaly between two satellites with the same initial conditions but experiencing different drag forces. This allows for the computation of the influence of a change in ballistic coefficient by a potential manoeuvre on a satellite orbit.\par
The necessary assumptions include a circular orbit over a spherical Earth, a non-rotating atmosphere and negligible change in the semi-major axis during the manoeuvre. 
The index $ r $ denotes an original reference orbit around Earth. A new perturbed orbit, indexed with $ n $, can be achieved by altering the ballistic coefficient of the satellite, \eg through a change in attitude. $ M_r $ and $ M_n $ further denote the respective mean anomalies of the orbits and 
\begin{equation}
	\phi(t) = M_r(t) - M_n(t)
\end{equation}
their difference. 
The mean motion $ n $ of a satellite is the time derivative of its mean anomaly, which results from Earth's gravitational parameter $ \mu_E $ and the orbit semi-major axis $ a $, it follows that
\begin{equation}
	\dot \phi_{rn} = n_r - n_n = \sqrt{\frac{\mu_E}{a_r^3}} - \sqrt{\frac{\mu_E}{a_n^3}}.
\end{equation}

Under the assumption that the change in semi-major axis during the manoeuvre is small compared to the original semi-major axis, \citet{Omar.2020} show that the first and second derivatives of the change in mean anomaly can be expressed as
\begin{align}
	\dot \phi_{rn} &= \frac{3\delta \sqrt{\mu_E} }{2 a_r^\frac{5}{2}}\\
	\ddot \phi_{rn} &= \frac{3 \sqrt{\mu_E} }{2 a_r^\frac{5}{2}} \dot{\Delta a}
\end{align}
where the change in semi-major axis is
\begin{equation}
	\Delta a = a_r-a_n.
\end{equation}

The time derivative of the orbital energy $ \dot E $ is directly linked to the magnitude of aerodynamic drag $ f_{D} $ and the satellite velocity $ v $ via the work-energy theorem:
\begin{align}
	E &= - \frac{\mu_E}{2a}\\
	\dot E &= \frac{\mu_E}{2a^2} \dot a = f_{D} v = -\frac{1}{2} \beta^* \rho v^3
	\label{eq:e_deriv}
\end{align}

Since the atmosphere is assumed to be non-rotating, the relative velocity equals the orbital velocity given by
\begin{equation}
	v = \sqrt{\frac{\mu_E}{a}}.
	\label{eq:orbital_velocity}
\end{equation}

Combining \cref{eq:e_deriv,eq:orbital_velocity} gives the time derivative of the semi-major axis as
\begin{equation}
	\dot a = -\beta^* \rho \sqrt{\mu_E a}.
\end{equation}

Finally, this leads to the time derivative of the change in semi-major axis:
\begin{equation}
	\Delta \dot a = \dot a_n - \dot a_r = - \rho \sqrt{\mu_E a} \left( \beta^*_n - \beta^*_r \right)
\end{equation}

Since $ \Delta a $ is small compared to the original semi-major axis, the semi-major axes can be assumed to be approximately equal $ a_r \approx a_n \approx a_0 $. Further, the density is assumed constant.
This yields an analytic expression for the second temporal derivative of $ \phi_{rn} $:
\begin{equation}
	\ddot \phi_{rn} = -\frac{3\rho \mu_E}{2a_0^2} \left( \beta^*_n - \beta^*_r \right) = -\frac{3\rho \mu_E}{2a_0^2} \Delta \beta^*
\end{equation}
This can be expressed from the perspective of the reference trajectory as
\begin{equation}
	\ddot \phi= \frac{3\rho \mu_E}{2a_0^2} \left( \beta^*_n - \beta^*_r  \right).
	\label{eq:omar_diff}
\end{equation}
By integration over time, one can deduce the built-up angular difference $ \phi$ between both trajectories. This angular separation can be converted into an in-track separation distance $ \Delta x $:
\begin{equation}
	\Delta x = \phi  a_0
	\label{eq:omar_dx}
\end{equation}\par

\begin{figure*}[h]
	\centering
%
%
%
%
	\includegraphics{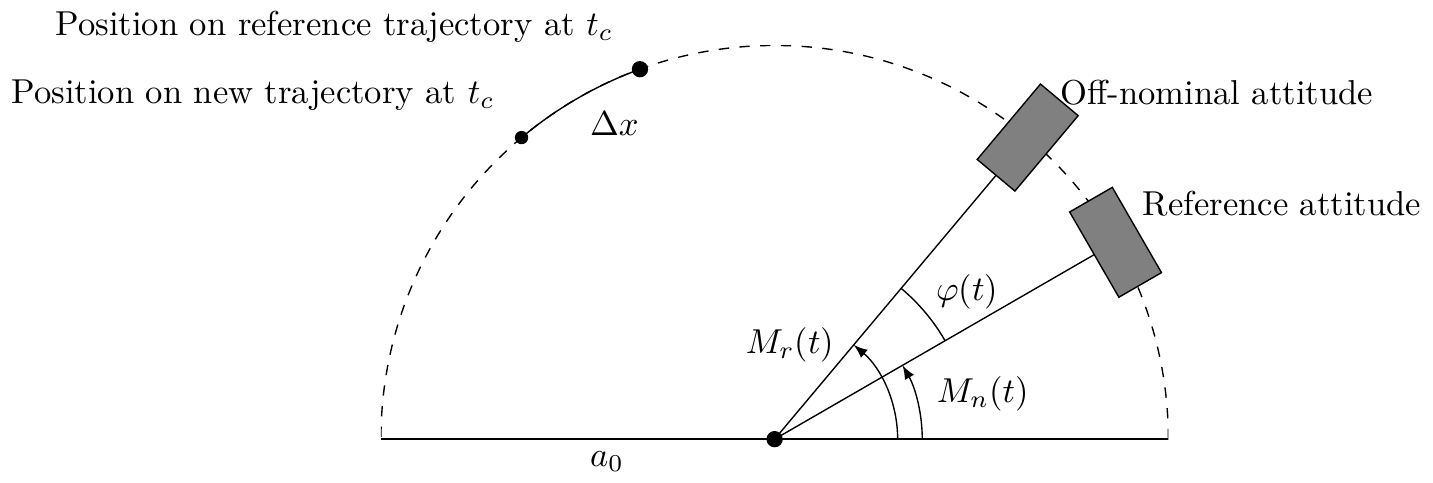}
	\caption{Concept of the collision avoidance manoeuvre using aerodynamic drag. The off-nominal attitude has an increased ballistic coefficient, leading to a positive in-track separation distance to the reference trajectory. It is to note, that the satellites started from equal initial conditions and the separation distance builds up over time.}
	\label{fig:manoeuvre_schematic}
\end{figure*}

An increase in ballistic coefficient ($\beta^*_n > \beta^*_r$) for a certain time leads to the satellite advancing in direction of flight from its original trajectory. This is schematically shown in \cref{fig:manoeuvre_schematic}. The satellite in off-nominal attitude has a higher ballistic coefficient, leading to a positive in-track separation distance compared to a flight in reference attitude at a later point in time $t_c$. Vice-versa, for a manoeuvre involving an attitude with a smaller-than-nominal ballistic coefficient the separation distance is negative.

\subsection{Conjunctions}
\label{sec:conjunctions}
A conjunction is a close encounter between a satellite and another object, posing an inherent threat because of the risk of a potential collision. This section describes how conjunction warnings are issued and how collision probabilities can be calculated.\par

\subsubsection{Conjunction Data Messages}
\label{sec:CDMs}
JSpOC provides services for satellite operators regarding conjunction events \cite{NASA.2020,SafetyHandbook.2020}. A proprietary catalogue of satellites and debris objects is continuously screened for close encounters with a miss distance below a defined threshold. When such a close encounter is identified, more thorough orbit and uncertainty analyses are performed and a standardized Conjunction Data Message (CDM) is sent to the satellite operators \cite{CCSDSCDM.2013}.
Each CDM contains extensive information about the encounter, \ie meta and orbital data of the objects, the predicted time of closest approach, predicted state vectors of both objects as well as estimated uncertainties and the models used for orbit propagation and respective parameters. The stated ballistic coefficient is of importance because it will be used as a reference value later on.

\subsubsection{Calculation of the collision probability}
\label{sec:pc_calculation}
Predictions about conjunctions are fraught with uncertainties. Therefore, a probability of collision associated with a close encounter is used as a main metric for risk assessment. \citet{Foster.1992} developed a method to calculate a collision probability between two objects for a given conjunction geometry by reducing the three-dimensional problem to a two-dimensional one. Many different implementations based on this original method have been derived since then \cite{NASA.2020, Foster.1992, Klinkrad.2006, Alfano.2005b} and JSpOC uses the method as well \cite{JSPOC.2016,CovarianceRealism.2016}. The FOSTER-1992 method makes the following assumptions:
\begin{itemize}
	\item The objects' relative velocity is so high that the encounter duration can be considered short and their motion can be assumed rectilinear, \ie with constant velocity and along a straight line, during the encounter.
	\item The uncertainties in the position components each follow a Gaussian distribution.
	\item Velocity uncertainties are neglected.
	\item The uncertainties are constant during the encounter, their values are the predicted values at the time of closest approach.
	\item The uncertainties in the primary and secondary object positions are uncorrelated and can thus be joined into the combined covariance by simple addition.
	\item Each object is represented by a sphere fully enclosing it. Its radius is called the hard body radius.
\end{itemize}

\begin{figure}[h]
	\centering
	\includegraphics[width=\linewidth]{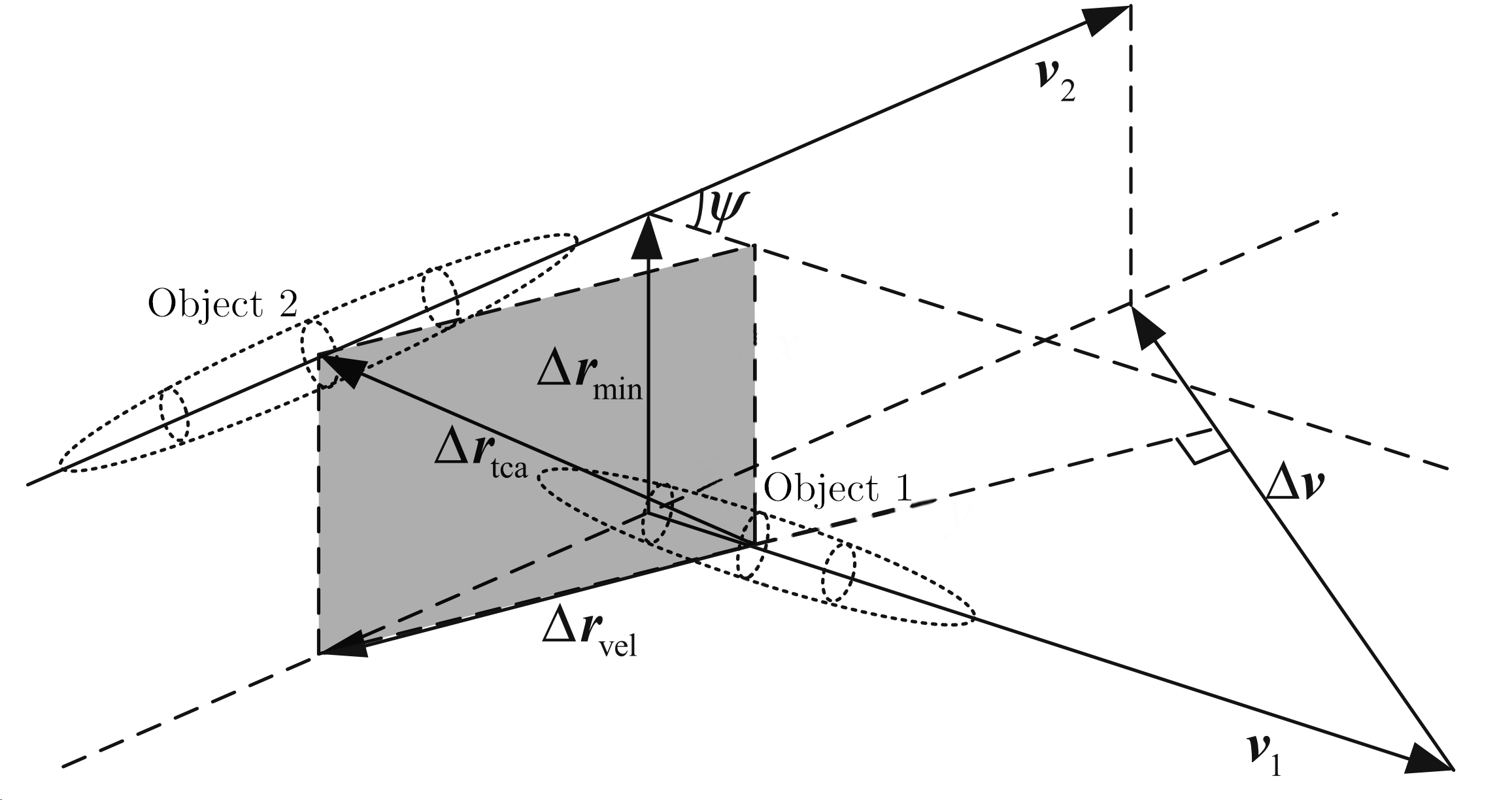}
	\caption{Illustration of the conjunction geometry at the time of closest approach with the conjunction plane in grey (adapted from \cite{Chen.2017}).}
	\label{fig:conjunction_geometry}
\end{figure}

\cref{fig:conjunction_geometry} depicts the geometry during a close encounter at the time of closest approach (TCA). The norm of the relative position vector between both objects at TCA $ \Delta\v{r}_{tca} = \v{r}_2 - \v{r}_1 $ is referred to as the miss distance. The plane normal to the relative velocity vector, which contains the relative position vector, is called the conjunction plane, or $ B $-plane. The relative position vector further defines the $ x $-axis of the conjunction frame.

Uncertainties in the satellite's and secondary object's positions $ \m{C}_{i,RTN} $ are specified as covariance matrices in the respective RTN-frames (radial, transversal, normal) \cite{Vallado}. Such a covariance matrix is symmetric and describes a three-dimensional ellipsoid indicating the $ 1\sigma $-environment of the position. The elements on the main diagonal represent the covariances $ \sigma_R^2,\sigma_T^2, \sigma_N^2 $, \ie the respective radial, transversal and out-of-plane components of the position uncertainty. The principle axes of the $ 1\sigma $-ellipsoid are defined by the eigenvectors of the covariance matrix and might deviate from the RTN-axes due to off-diagonal elements not being equal to zero.\par

\citet{Foster.1992} transform the two objects' covariances to the ECEF-frame (Earth-centred Earth-fixed) and add them to form one combined covariance $ \m{C}_{tot,ECEF} $:
\begin{equation}
	\m{C}_{tot,ECEF} = \m{C}_{1,ECEF} + \m{C}_{2,ECEF}
\end{equation}
The combined covariance is then assigned to the secondary object. On the other hand, the two objects' hard body radii are summed as well and assigned to the primary object, yielding the collision sphere with radius $ R_C $. Therefore, a collision occurs if the distance between both objects at $ \tca $ is smaller than the radius of the collision sphere, \ie the collision sphere is intersected. Due to the assumption of rectilinear motion, it is possible to convert the three-dimensional problem to a two-dimensional one by projecting the combined covariance ellipsoid and the collision sphere onto the conjunction plane. The resulting covariance ellipse on the conjunction plane is characterized by the projected combined covariance matrix:
\begin{equation}
	\m{C}_{tot,B} =  \m{R}_{B,ECEF} \m{C}_{tot,ECEF} \transp{\m{R}_{B,ECEF}}
\end{equation}
where $\m{R}_{B,ECEF}$ is the transformation matrix from ECEF-frame to the conjunction plane.
The projected covariance $ \m{C}_{tot,B} $ represents a Gaussian probability distribution in the $ B $-plane. By integrating the probability density function over the collision area $ A_C $ (\ie the projected collision sphere), a probability of collision $ P_c $ can be estimated:
\begin{equation}
	P_c = \frac{1}{2\pi \sqrt{\det(\m{C}_{tot,B})}} \int_{-R_C}^{R_C} \int_{-\sqrt{R_C^2-x_B^2}}^{\sqrt{R_C^2-x_B^2}} \exp(-A_B)\D y_B \D x_B
	\label{eq:pc_foster}
\end{equation}
\begin{equation*}
	\text{with:\quad} A_B = \frac12 \transp{\Delta \v{r}_B} \m{C}_{tot,B}^{-1} \Delta \v{r}_B
\end{equation*}

The collision probability is significantly influenced by shape and size of the covariance ellipsoid. \citet{Alfano.2005} developed a method to determine the maximum collision probability based on the aspect ratio $ \lambda = \frac{\sigma_{B,1}}{\sigma_{B,2}}$ between the semi-major and semi-minor axes of the projected covariance ellipse \cite{Alfano.2005,Mishne.2017}. 
The maximum collision probability results from
\begin{equation}
	P_{c,max,alf} = \frac{R_C^2}{2\lambda\sigma^2} \exp{\left( -\frac12 \left( \frac{\Delta r_{tca}}{\lambda\sigma} \right)^2 \right)}
	\label{eq:pc_max_alfano}
\end{equation}
\begin{equation*}
	\text{with\quad} \sigma=\frac{\Delta r_{tca}}{\sqrt{2}\lambda}.
\end{equation*}
This metric includes information about the eccentricity of the combined covariance matrix. 

\section{Analysis tool}
\label{ch:methodology}
In the following section, the methodology behind the analysis tool for collision avoidance using aerodynamic drag is explained. 

\subsection{The manoeuvre}
\label{sec:manoeuvre}
In \cref{sec:sep_distance}, an analytic equation was presented which allows the calculation of a relative in-track separation distance by the variation of a satellite's ballistic coefficient relative to a reference value. During conjunction screening and the orbit determination process, the JSpOC determines a mean ballistic coefficient from observations, which is used for propagation as well as conjunction analysis and will be considered the reference value. By deliberately taking an (off-nominal) attitude with a ballistic coefficient different to the reference value until the predicted TCA, a separation distance can be built up relative to JSpOC's predicted reference trajectory. 

For satellites changing their ballistic coefficient via changes in attitude, manoeuvring constraints might be necessary. A common example are attitudes which cause the solar arrays to point away from the Sun and thus decrease the power generated by them. Such attitudes can only be held as long as battery power is sufficient. Other constraints might include pointing requirements imposed by mission objectives etc.
To be able to consider these constraints in the manoeuvre analysis, the manoeuvring time is divided into sections. Each section consists of two parts of specified duration $ t_1 $ and $ t_2 $, respectively. During the first part, the commanded manoeuvring attitude is attained. During the second part, a predefined attitude is implemented to meet the constraints.\par

\begin{figure*}[htb!]
	\centering
	\includegraphics{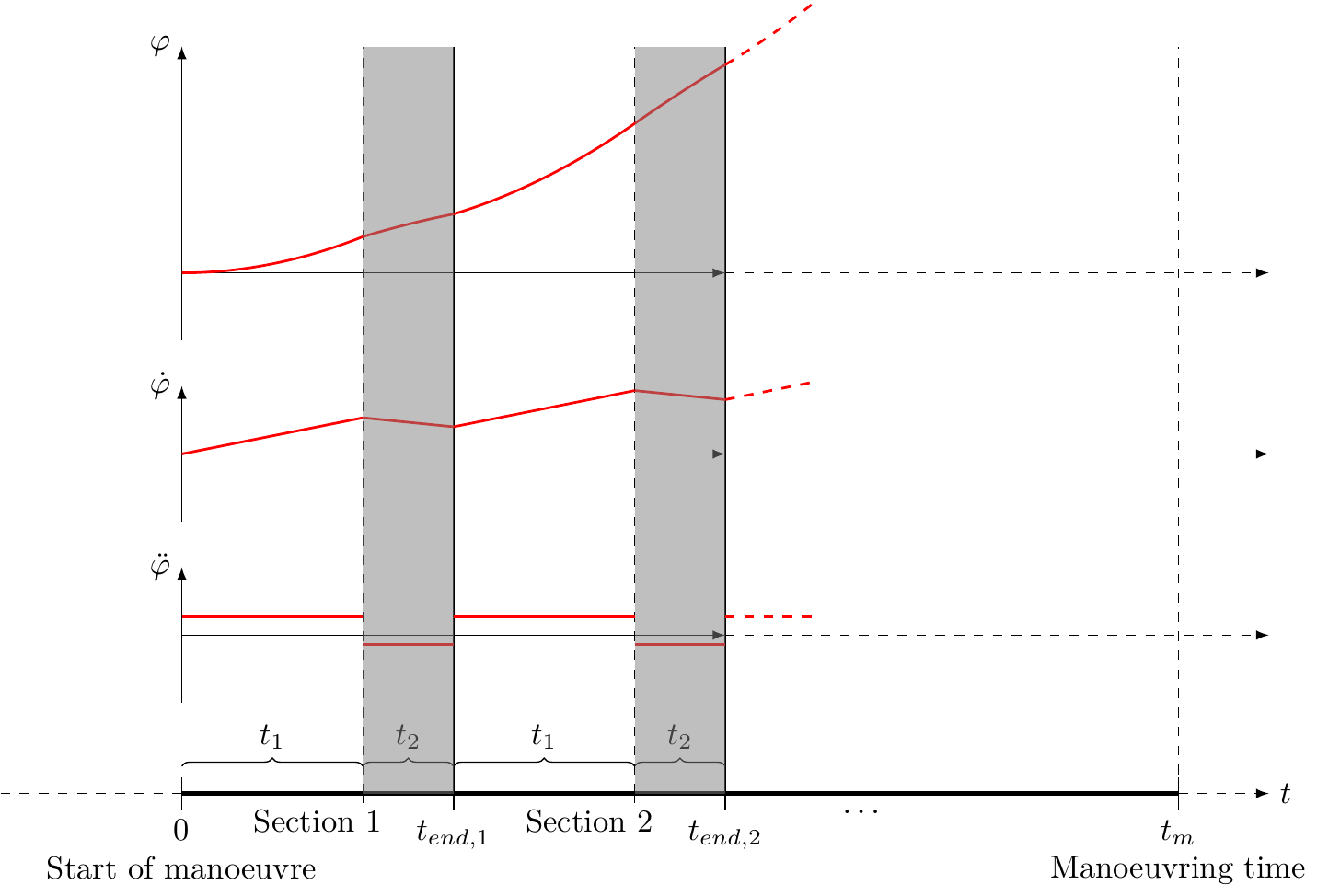}
	\caption{Exemplary consideration of operational constraints. Considered is a manoeuvre with a commanded inverse ballistic coefficient $ >\beta^*_{ref} $ and additional phases (grey), in which constraints dictate a flight with $ <\beta^*_{ref} $. The respective evolution of the angular separation $ \phi $ between the reference and the perturbed trajectory as well as its derivatives are shown.}
	\label{fig:manoeuvre_timeline2}
\end{figure*}

\cref{fig:manoeuvre_timeline2} shows an exemplary timeline of such a manoeuvre and the respective courses of the angular separation $ \phi $ as well as its first and second derivatives. While the second derivative is directly proportional to the change in inverse ballistic coefficient and therefore constant in phases with constant attitude, the first derivative is a piece-wise linear function representing an angular motion relative to the reference trajectory.
In this example, the inverse ballistic coefficient in the commanded manoeuvring attitude is higher than the reference value. Therefore, the angular separation $ \phi $ grows quadratically while the satellite is in manoeuvring attitude. During the constrained charging phases, the inverse ballistic coefficient is $ \beta^*_{constr}<\beta^*_{ref} $, here. The negative value of the change in inverse ballistic coefficient and hence $ \ddot \phi $ causes the first derivative of the angular separation to decrease. $ \phi $ itself still increases, as long as the relative angular motion $ \dot \phi > \SI{0}{\per \second} $. The separation distance is directly proportional to $ \phi $ and therefore shows the same behaviour. The respective courses vary depending on the actual ballistic coefficients in the different phases.

This already shows that a manoeuvre is more effective the earlier it is implemented. Due to the build-up of relative angular motion $ \dot \phi $, the angular separation can be increased despite negative values of $ \ddot \phi $ in later phases. These might be caused by phases with an inverse ballistic coefficient lower than the reference value. In case of a return to the reference attitude, in which $ \ddot \phi = \SI{0}{\per \square\second} $, $ \phi $ will still grow linearly over time because of a constant relative angular motion.

\subsection{The tool}
\label{sec:tool}
\begin{figure}[H]
	\centering
%
%
	\includegraphics{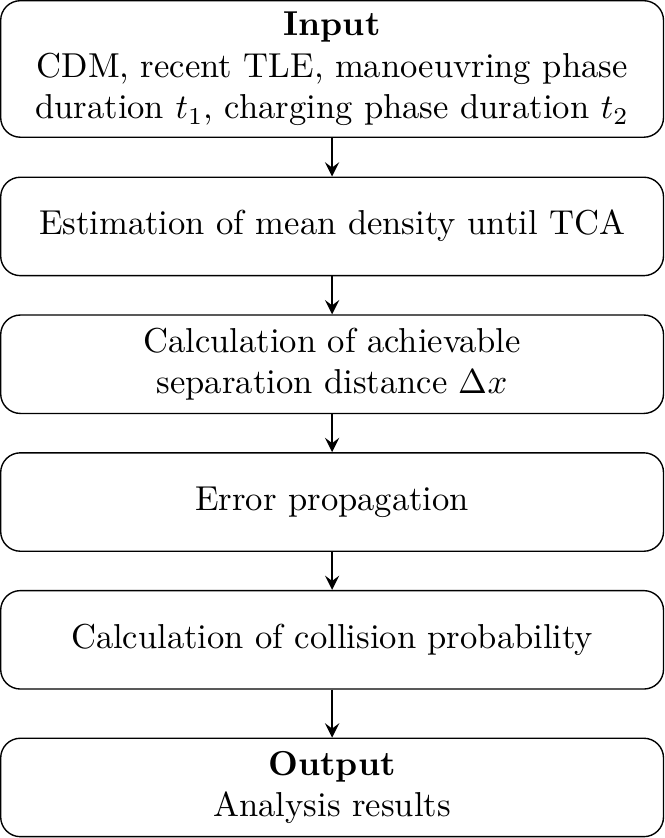}
	\caption{Flowchart of the developed analysis tool.}
	\label{fig:tool_flowchart}
\end{figure}

The tool is based upon \cref{eq:omar_diff} by \citet{Omar.2020}. \cref{fig:tool_flowchart} shows its  basic concept. Using only a CDM and pointing constraints as user inputs, the influence of a potential manoeuvre on the collision probability can be analysed.\par

The tool is programmed in \code{python} and can be operated via a Jupyter Notebook. The tool makes use of the \code{pyatmos} package for atmosphere modelling \cite{pyatmos.2021}, which has been slightly adapted. Sgp4-propagation algorithms are implemented with a package developed at IRS, which itself is based on the \code{skyfield} package \cite{skyfield.2022}.   

\subsubsection{Loading input data}
A CDM is issued by JSpOC when a close encounter is detected and contains all the necessary information about a conjunction event. It serves as primary input for the tool. Besides the predicted conjunction geometry and the associated uncertainties, especially the satellite's reference inverse ballistic coefficient $ \beta^*_{ref} $ is of interest to examine a potential avoidance manoeuvre.
For the propagation of the satellite position in the next step, a current two-line element set (TLE) is loaded. This is also used to provide the satellite's semi-major axis $ a_0 $. The TLE can be loaded from \url{space-track.org} \cite{space-track}, where the 18th Space Defense Squadron publishes TLEs obtained through own observations.

\subsubsection{Density estimation}
The achievable separation distance during a specified interval depends on the average density the satellite will encounter. To estimate the average atmospheric density during the manoeuvre, an atmospheric model considering is evaluated at a number of points along the reference trajectory. The previously obtained TLE is propagated for this using the sgp4-algorithm. 
The deviation in position due to the manoeuvre is considered negligible here, since it is small and only occurs in the horizontal plane and thus only minimally affects the density output.
The averaged density $ \bar \rho $ then results from the evaluation of the chosen atmospheric model at $ n $ evenly-distributed time steps $ t_i $:
\begin{equation}
	\bar \rho = \frac1n \sum_{i=1}^{n} \rho(\v{r}_i, t_i, \v{K}_i)
\end{equation}
where $ \v{r}_i $ denotes the satellite's position and $ \v{K}_i $ the space-weather indices for the atmosphere model at time step $ t_i $. 
The NRLMSISE-00 and JB-2008 density models are implemented and can be chosen for analysis. 
The \code{pyatmos} package retrieves recent space-weather data from \citet{SET}'s website. The indices for solar activity are updated on a daily basis, whereas the geomagnetic index is updated every three hours.

\subsubsection{Calculation of achievable separation distance}
In principle, a manoeuvre can be performed by taking any specified attitude. The value for the respective inverse ballistic coefficient has to be provided.
For clarity, the manoeuvring inverse ballistic coefficient will be referred to as $ \beta^* $ in the following, the reference value as $ \beta^*_{ref} $ and the value for the constraint phases as $ \beta^*_{constr} $.

At the start of a manoeuvre, the separation distance as well as its derivative equal zero: $ \dot \phi(\SI{0}{\second})=\phi(\SI{0}{\second})=\SI{0}{\per \second} $. For section $ i $, the separation distance at the end of the phase can then be calculated as follows:
\begin{align}
	\ddot \phi_{1,i} = &\frac{3 \bar \rho \mu_E}{2a_0^2} \left( \beta^* - \beta^*_{ref} \right)\\
	\ddot \phi_{2,i} = &\frac{3\bar \rho \mu_E}{2a_0^2} \left( \beta^*_{constr} - \beta^*_{ref} \right)
\end{align}
The second derivative $ \ddot \phi $ of the angular difference between the mean anomaly of the reference and manoeuvring trajectory is constant within each of the two parts of the section.

The first derivative $ \dot \phi $ is a piece-wise linear function of gradient $ \ddot \phi $. It follows through integration over time:
\begin{align}
	\begin{split}
		\dot \phi_{1,i}(t) = 	&\ddot \phi_{1,i} \left[t-t_{end,i-1}\right]\\
		&+ \dot \phi_{2,i-1}(t_{end,i-1})
	\end{split}\\[2ex]
	\begin{split}
		\dot \phi_{2,i}(t) = 	&\ddot \phi_{2,i} \left[t-(t_{end,i-1}+t_1)\right]\\
		&+ \dot \phi_{1,i}(t_{end,i-1}+t_1)
	\end{split}
\end{align}
Finally, the angular separation of the mean anomalies can be obtained via another integration over time, yielding
\begin{align}
	\begin{split}
		\phi_{1,i}(t) = &\frac12 \ddot \phi_{1,i} \left[t-t_{end,i-1}\right]^2\\
		&+ \dot \phi_{2,i-1}(t_{end,i-1}) \left[t-t_{end,i-1}\right]\\
		&+ \phi_{2,i-1}(t_{end,i-1})
	\end{split}\\[2ex]
	\begin{split}
		\phi_{2,i}(t) = &\frac12 \ddot \phi_{2,i} \left[t-(t_{end,i-1}+t_1)\right]^2\\
		&+ \dot \phi_{1,i}(t_{end,i-1}+t_1) \left[t-(t_{end,i-1}+t_1)\right]\\
		&+ \phi_{1,i}(t_{end,i-1}+t_1).
	\end{split}
\end{align}
$ \phi_{2}(t_m) $ at the end of the last section is the desired separation, which can be translated into a separation distance via \cref{eq:omar_dx}.

If no constraint phases are considered, the resulting separation distance follows as:
\begin{align}
	\Delta x = -\frac12 \ddot\phi t_m^2 = \frac{3\bar \rho \mu_E}{4a_0} \Delta \beta^* t_m^2
	\label{eq:deltax_min}\\
	\text{with\quad} \Delta \beta^* = \left( \beta^* - \beta^*_{ref} \right)
\end{align}
$ \Delta \beta^* $ is the change in inverse ballistic coefficient in the commanded attitude compared to the reference value.

\subsubsection{Error propagation}
The influence of uncertainties in the parameters of \cref{eq:deltax_min} may in a first step be evaluated by applying a Gaussian error propagation \cite{Papula.2011}. The standard deviations of the averaged density, the semi-major axis, the difference in inverse ballistic coefficient and the manoeuvring time are denoted as $ \sigma_{\bar \rho} $, $ \sigma_{a_0} $, $ \sigma_{\Delta \beta^*} $, and $ \sigma_{t_m} $, respectively. The resulting standard deviation of the separation distance can be calculated by
\begin{equation}
	\begin{split}
		\sigma_{\Delta x} = 
		&\sqrt{	\left(\frac{\partial \Delta x}{\partial \bar \rho}\sigma_{\bar \rho}\right)^2
			+ \left(\frac{\partial \Delta x}{\partial a_0}\sigma_{a_0}\right)^2}\\
		&\overline{+ \left(\frac{\partial \Delta x}{\partial \Delta \beta^*}\sigma_{\Delta \beta^*}\right)^2
			+ \left(\frac{\partial \Delta x}{\partial t_m}\sigma_{t_m}\right)^2}.
		\label{eq:error_propagation}
	\end{split}
\end{equation}%
The partial derivatives with respect to one parameter are to be evaluated with the other parameters set to their nominal value. They are given by:
\begin{align}
	\frac{\partial \Delta x}{\partial \bar \rho} &= \frac{3 \mu_E}{4a_0} \Delta \beta^* t_m^2 \label{eq:error_propagation_rho}\\
	\frac{\partial \Delta x}{\partial a_0} &= -\frac{3\bar \rho \mu_E}{4a_0^2} \Delta \beta^* t_m^2 \label{eq:error_propagation_sma}\\
	\frac{\partial \Delta x}{\partial \Delta \beta^*} &= \frac{3 \bar \rho \mu_E}{4a_0} t_m^2 \label{eq:error_propagation_CB}\\
	\frac{\partial \Delta x}{\partial t_m} &= \frac{3 \bar \rho \mu_E}{2a_0} \Delta \beta^*  t_m \label{eq:error_propagation_tc}
\end{align}
The uncertainties in the parameters may be described by an uncertainty level $ s \ge 0 $, so that $ \sigma_p=s_p p $ is the respective standard deviation of any parameter $ p $. It follows that uncertainties of a given uncertainty level in any of the first three parameters affect the standard deviation of the achievable separation distance in the same way. The manoeuvring time, however, has an impact that is higher by a factor of $ 2 $.

The uncertainty in the separation distance leads to an additional uncertainty $ \sigma_{\Delta x} $ of the satellite position in the direction of flight. With the assumption that the additional uncertainty is uncorrelated to the uncertainty stated in the CDM $ \sigma_T $, these two can be added:
\begin{equation}
	\sigma_T' = \sigma_{T} + \sigma_{\Delta x}
\end{equation}
$ \sigma_T' $ is the updated transverse position uncertainty, which will be used to calculate the collision probability in the next step. The assumption of the uncertainties being uncorrelated is a first approach and needs to be further investigated in future work.

\subsubsection{Calculation of collision probability}
An avoidance manoeuvre changes the conjunction geometry. Therefore, a new point of closest approach has to be determined considering the perturbed trajectory of the satellite. Afterwards, the collision probability is calculated with the Foster-1992 method, as described in \cref{sec:pc_calculation}. All assumptions mentioned there apply.\par
The position of the manoeuvring satellite at the original TCA $ t_{tca} $ can be determined by regarding the built-up separation distance:
\begin{equation}
	\v{r}_{1}' \left(t_{tca}\right) = \v{r}_{1,tca} + \Delta x \frac{\v{v}_{1,tca}}{\left| \v{v}_{1,tca} \right|}
\end{equation}
Based on this, a new time of closest approach $ t_{tca}' $ may be determined by finding the minimum distance between the objects over time:
\begin{equation}
	\begin{split}
		t_{tca}' = t_{tca}+\Delta t \\
		\Rightarrow \left| \Delta \v{r} \right|
		&= \left| \v{r}_2 - \v{r}_1' \right|\\
		&= \left| \v{r}_{2,tca} + \Delta t \v{v}_{2,tca} - \v{r}_{1,tca}' - \Delta t \v{v}_{1,tca} \right|\\ &\text{\quad is minimum}
	\end{split}
\end{equation}

The elements of the covariance matrix describing the position uncertainty of the satellite are updated according to
\begin{equation}
	C_{RTN,1,{i,j}}' = 
	\begin{cases}
		\sigma_T' = \left(\sigma_{T} + \sigma_{\Delta x}\right)^2 &, \ i=j=1\\
		C_{RTN,1,{i,j}} &, \ \text{else}
	\end{cases}.
\end{equation}
Together with the updated time and point of closest approach, a probability of collision may be calculated employing \cref{eq:pc_foster} for the different manoeuvring attitudes and depending on the manoeuvring time.

The tool's computed collision probabilities were verified using exemplary conjunctions from literature \cite{Klinkrad.2006,Chen.2017,CARA}. Notably, this is not the value determined by JSpOC, since they perform further scaling of the covariance matrices before determining the collision probability, resulting in different collision probabilities stated in the CDM \cite{JSPOC.2016, CovarianceRealism.2016}.
\section{Results}
\label{sec:results_discussion}
This chapter presents the results of analyses performed with the tool on exemplary close encounters of the \flp. Achievable separation distances are investigated before the influence of parameter uncertainties on the collision probability is analysed. Solar and geomagnetic data for the different activity levels are defined as in \cref{tab:activity_levels}, the used atmosphere model is NRLMSISE-00. The aerodynamic data of the \flp are defined in \cref{tab:aero_results} and determined via interpolation depending on the space weather data. Only the minimum and maximum drag attitudes of the \flp are taken into account, since they maximize the difference in ballistic coefficient compared to the reference value thus also maximizing the achievable separation distance. For the constraint phases, the Nadir-pointing attitude is assumed.

\subsection{Achievable separation distance}
\label{sec:ana_achievable_separation_distance}
In this section, the maximum separation distance of the \flp achievable via a change in ballistic coefficient is examined. 
Its reference trajectory is based on a CDM received for a close encounter on April 7, 2022, while different levels of solar and geomagnetic activity are assumed. The reference inverse ballistic coefficient is $ \beta^*_{ref} = \SI{0.01794}{\square\meter\per\kilogram}$, which is the average of recent CDMs.
The density is evaluated and averaged over one orbital period of the reference orbit. The resulting values are 
$ \bar \rho = \SI{1.158e-14}{\kilogram\per\cubic\meter} $ for low, 
$ \bar \rho = \SI{1.650e-13}{\kilogram\per\cubic\meter} $ for moderate and 
$ \bar \rho = \SI{1.020e-12}{\kilogram\per\cubic\meter} $ for high activity.

\begin{figure*}[p!]
	\centering
	\includegraphics{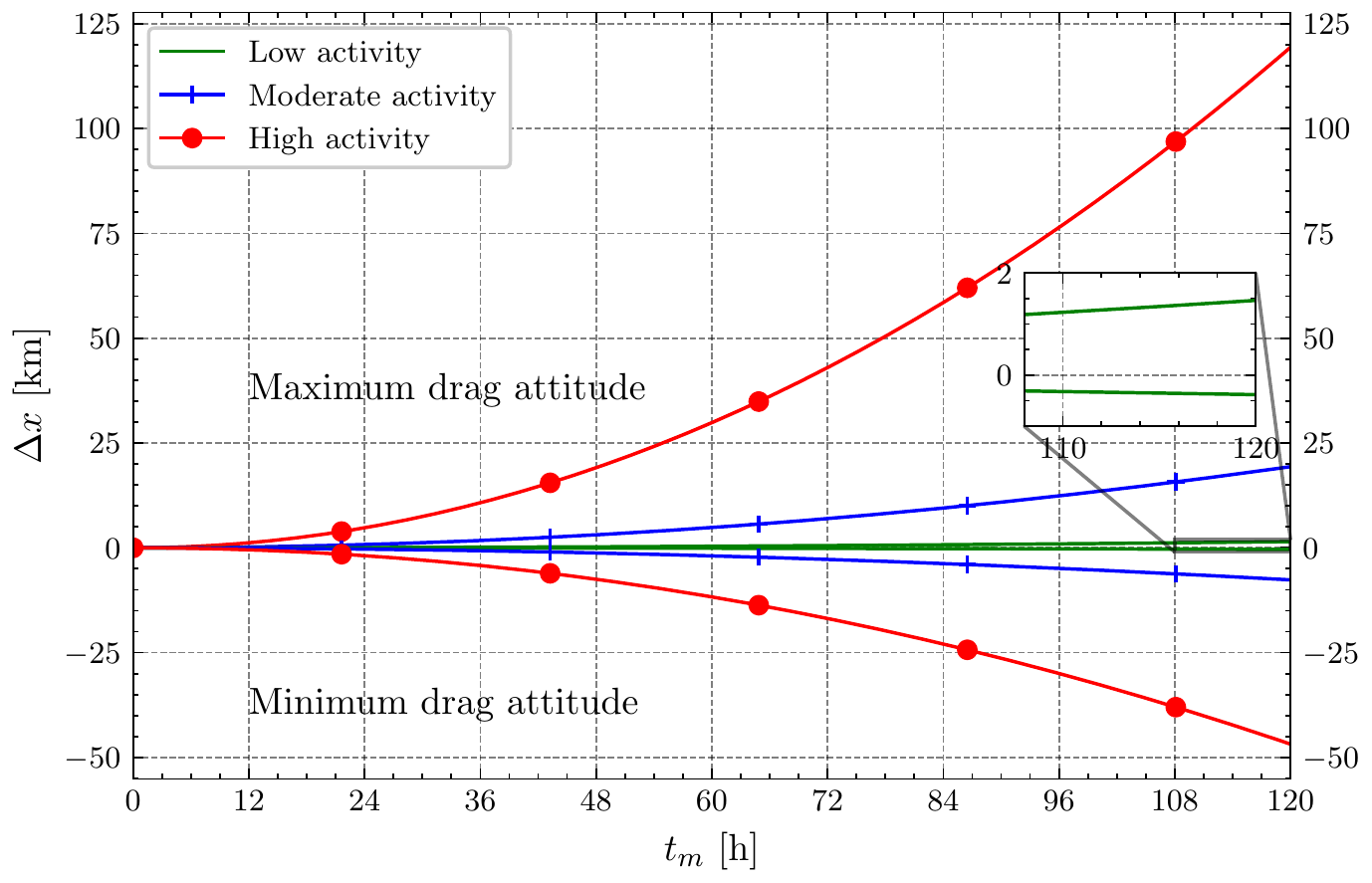}
	\caption{Achievable separation distance of the \flp for different levels of solar and geomagnetic activity.}
	\label{fig:analysis_sep_dist}
\end{figure*}

The tool is used to analytically estimate the achievable separation distance for a given manoeuvring time. The results are presented in \cref{fig:analysis_sep_dist}. The positive separation distances are established by a flight with increased inverse ballistic coefficient. Consequently, a decreased inverse ballistic coefficient allows for the build-up of negative separation distances. Further, the achievable separation distance grows linearly with increasing atmospheric density, superimposed on it is the smaller influence of the density on the ballistic coefficient (\cf \cref{eq:omar_dx}). For low solar and geomagnetic activity, the resulting low density causes the separation distance to turn out negligibly small at only $ \SI{1.465}{\kilo\meter} $ for a flight in maximum drag attitude and even smaller for a flight with minimum drag. At moderate activity, the separation distances are noticeably higher at $ \SI{19.35}{\kilo\meter} $ and  $ \SI{-7.647}{\kilo\meter} $, respectively. For a high level of activity, the separation distances grow up to $ \SI{119.4}{\kilo\meter} $ and  $ \SI{-46.81}{\kilo\meter} $.\par

\subsection{Influence of manoeuvring constraints}
\begin{figure*}[p!]
	\centering
	\includegraphics{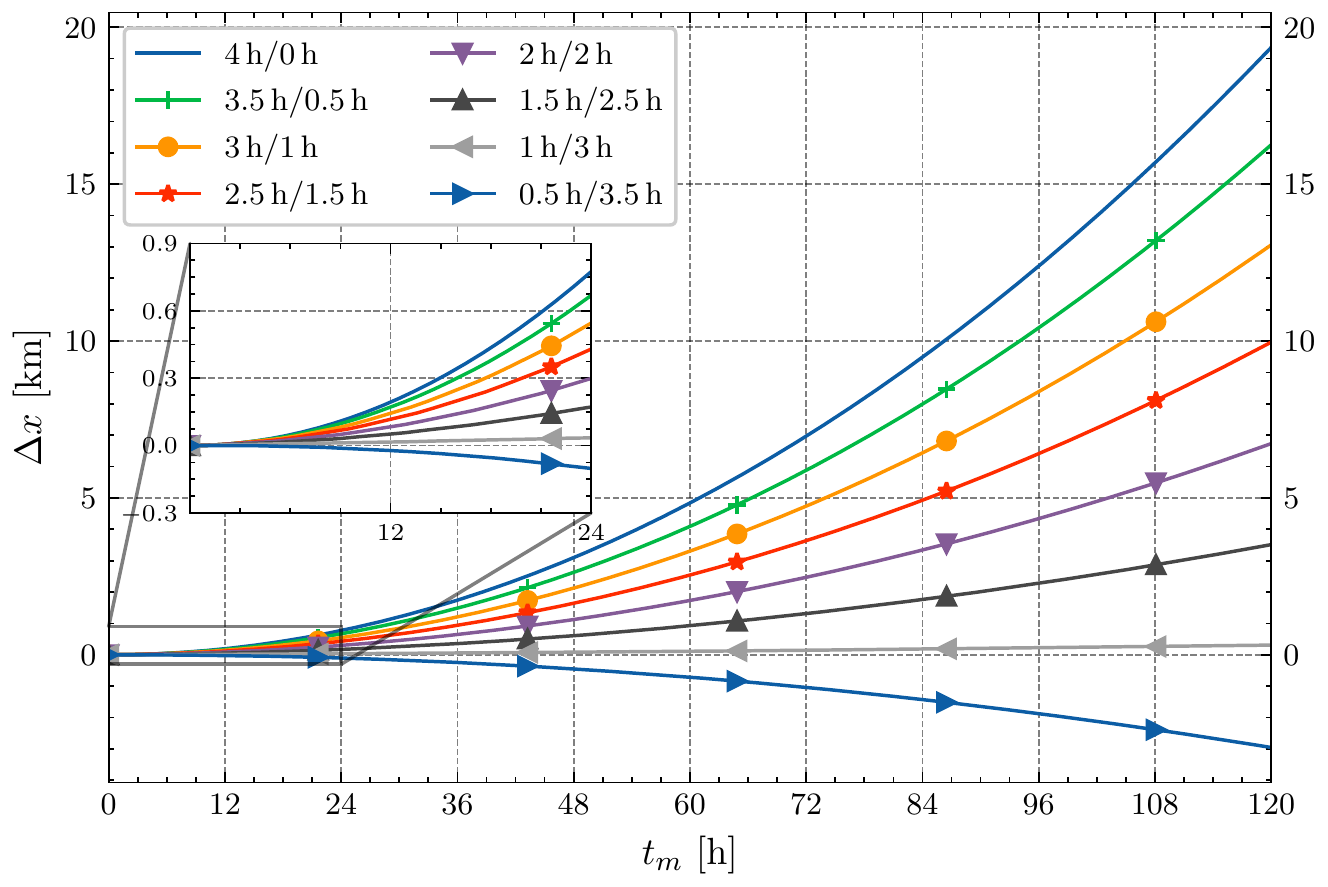}
	\caption{Achievable separation distance of the \flp for different constraints and moderate solar and geomagnetic activity. The constraints consider a manoeuvring and charging phase of variable duration, \eg $ \SI{3}{\hour} $/$ \SI{1}{\hour} $ is a flight in maximum drag attitude for $ \SI{3}{\hour} $, followed by a charging phase of $ \SI{1}{\hour} $.}
	\label{fig:analysis_sep_dist_constrained}
\end{figure*}

Next, the effect of additional constraints on the achievable separation distance shall be evaluated. The scenario is the flight in maximum drag attitude for the manoeuvre mentioned before at moderate solar and geomagnetic activity. Additionally, phases in which the \flp points its solar panels to the Sun to re-charge is batteries are now taken into account. Therefore, the manoeuvring time is divided into individual sections of $ \SI{4}{\hour} $ duration consisting of two parts. In the first phase, the satellite takes the commanded attitude, while in the second an offset Nadir-pointing is performed, which results in an attitude suitable for charging. It results from a $ \SI{90}{\degree} $ rotation of the satellite around the direction of flight starting from the Nadir-pointing attitude. The duration of the two phases is varied while adding up to $ \SI{4}{\hour} $. \cref{fig:analysis_sep_dist_constrained} shows the resulting achievable separation distances depending on the available time. If no charging phase is carried out ($ \SI{4}{\hour} $/$ \SI{0}{\hour} $), the separation distance is equal to the one for moderate activity and maximum drag in \cref{fig:analysis_sep_dist}. For longer charging phases, the achievable separation distance decreases. It does, however, still develop monotonously. For a manoeuvre involving sections of $ \SI{1}{\hour} $ flight in maximum drag attitude followed by $ \SI{3}{\hour} $ of charging, the effects of both phases almost cancel out. If the charging phase is commanded even longer in relation to the first phase, the separation distance becomes negative, due to the ballistic coefficient in Nadir-pointing being smaller than the reference ballistic coefficient. Within the first $ \SI{12}{\hour} $, all manoeuvres' separation distances are below $ \SI{300}{\meter} $. After $ \SI{24}{\hour} $, the $ \SI{2}{\hour} $/$ \SI{2}{\hour} $ split manoeuvre achieves a separation distance of $ \sim \SI{300}{\meter} $. The quadratic course of the separation distance then leads to increasingly higher values.

\subsection{Influence of relative in-track position at TCA}
\label{sec:ana_conjunction_geometry}
\begin{figure}[h!]
	\centering
	\begin{subfigure}{\linewidth}
		\centering
		\includegraphics{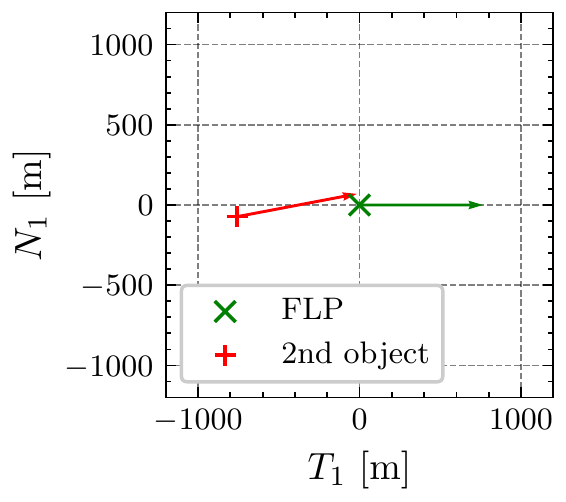}
		\caption{Encounter A. The radial component of the relative position is $ \SI{0.1}{\meter} $.}
		\label{fig:conjunction_geometries_a}
	\end{subfigure}\\
	\begin{subfigure}{\linewidth}
		\centering
		\includegraphics{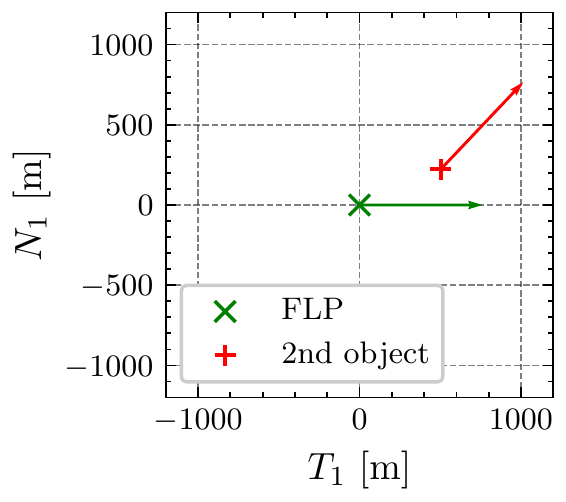}
		\caption{Encounter B. The radial component of the relative position is $ \SI{-61}{\meter} $.}
		\label{fig:conjunction_geometries_b}
	\end{subfigure}
	\caption{Predicted conjunction geometries at TCA projected onto the local horizontal plane of the \flp. Velocities are scaled by a factor of $ 0.1 $.}
	\label{fig:conjunction_geometries}
\end{figure}

To study the effects of the predicted conjunction geometry on possible manoeuvres, two past close encounters are compared in the following section. The first is the previous close encounter on April 7, 2022, (encounter A), while the second is a close encounter on March 30, 2022, (encounter B). The two conjunctions differ with regards to the in-track component of the predicted relative position, which is negative for encounter A and positive for encounter B. This is visible in \cref{fig:conjunction_geometries}.

\begin{figure}[h]
	\centering
	\begin{subfigure}{\linewidth}
		\centering
		\resizebox{\linewidth}{!}{\includegraphics{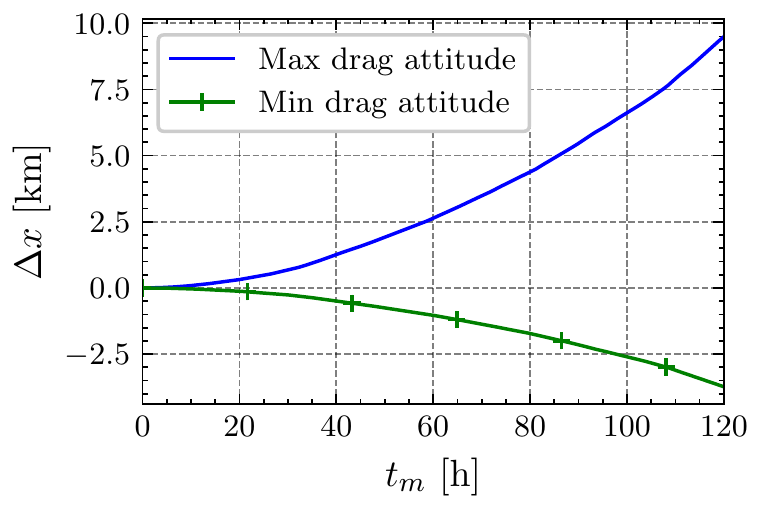}}
		\caption{Encounter A.}
		\label{fig:sep_distances_a}
	\end{subfigure}\\
	\begin{subfigure}{\linewidth}
		\centering
		\resizebox{\linewidth}{!}{\includegraphics{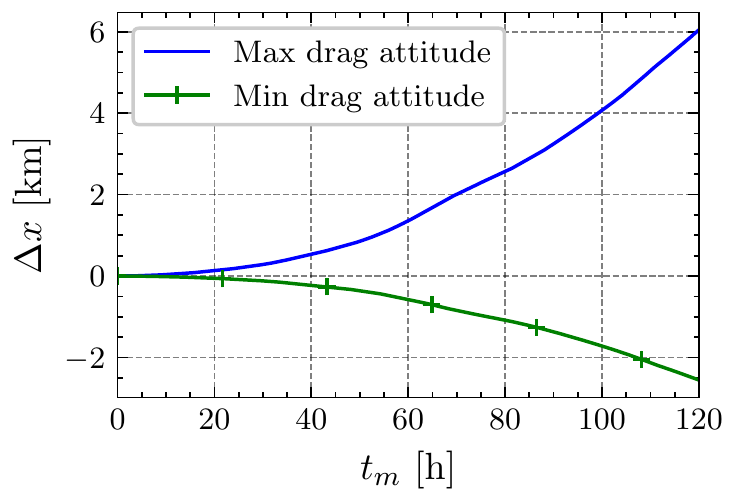}}
		\caption{Encounter B.}
		\label{fig:sep_distances_b}
	\end{subfigure}
	\caption{Achievable separation distances for both encounters depending on manoeuvring time.}
	\label{fig:sep_distances}
\end{figure}

A maximum manoeuvring time of five days is assumed for both scenarios. The resulting achievable separation distances are presented in \cref{fig:sep_distances}. For both encounters, the achievable separation distance in maximum drag attitude is higher.

\begin{figure}[h]
	\centering
	\begin{subfigure}{\linewidth}
		\centering
		\resizebox{\linewidth}{!}{\includegraphics{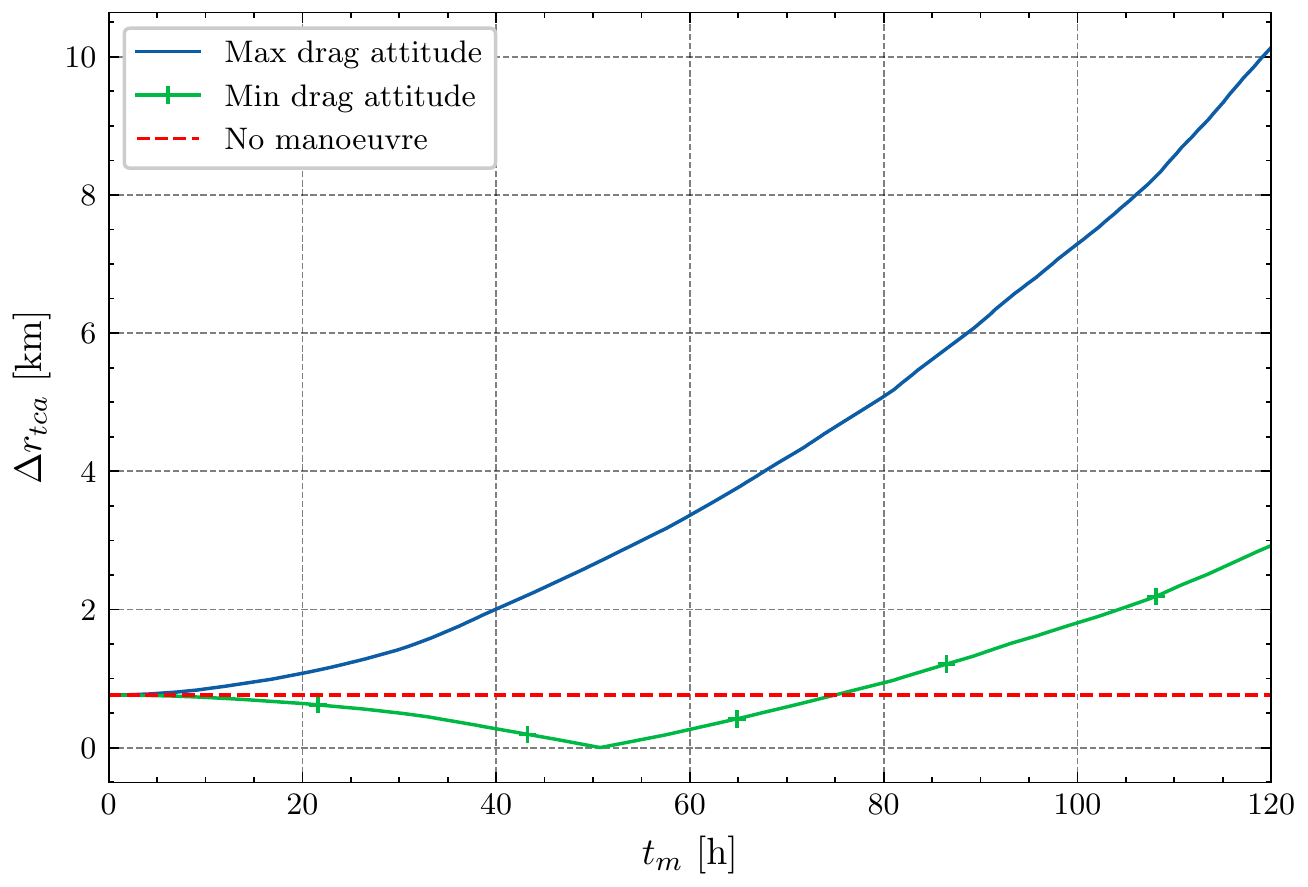}}
		\caption{Encounter A.}
		\label{fig:miss_distances_a}
	\end{subfigure}\\
	\begin{subfigure}{\linewidth}
		\centering
		\resizebox{\linewidth}{!}{\includegraphics{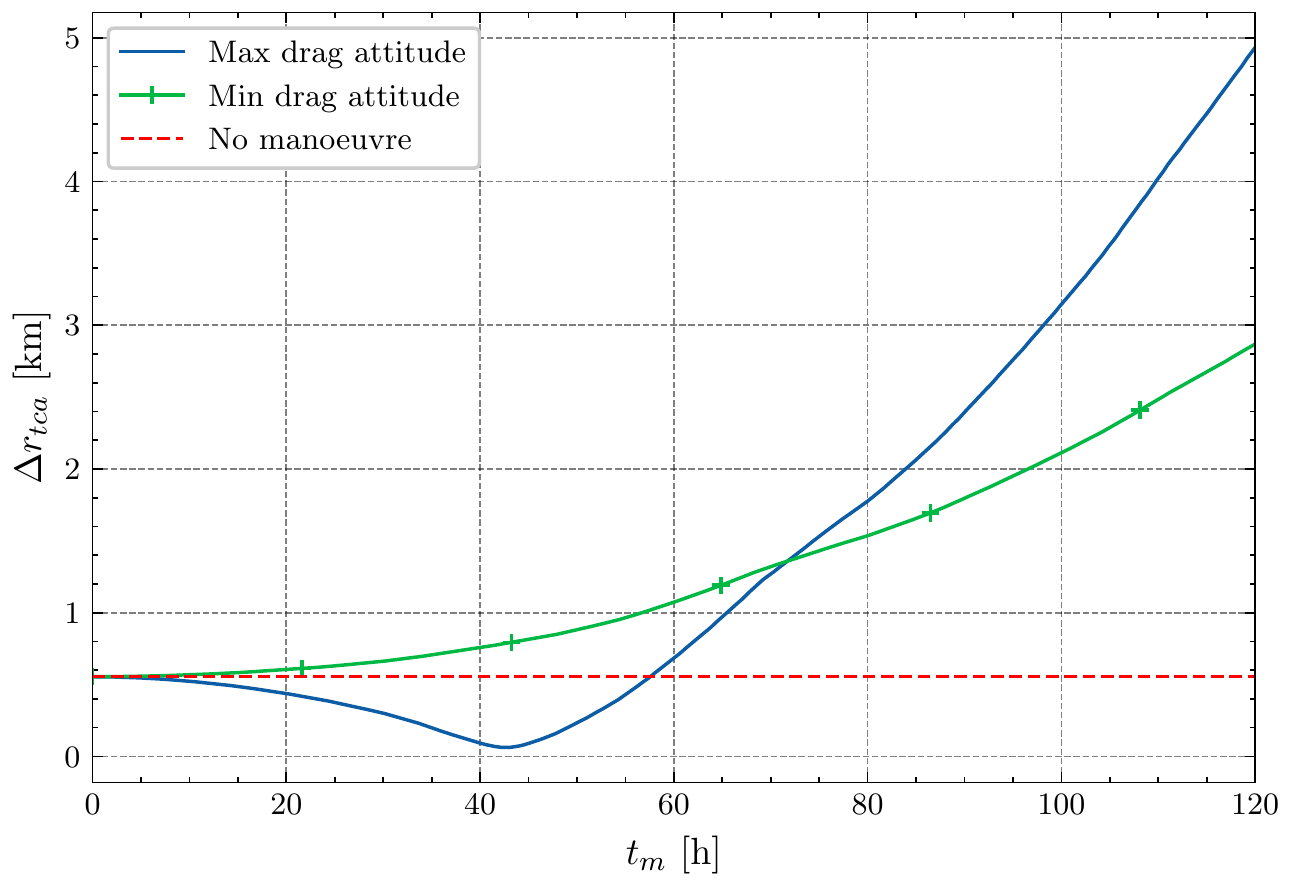}}
		\caption{Encounter B.}
		\label{fig:miss_distances_b}
	\end{subfigure}
	\caption{Miss distance depending on manoeuvring time.}
	\label{fig:miss_distances}
\end{figure}

The miss distances at TCA shall be evaluated more carefully and are visualized in \cref{fig:miss_distances}. For the first encounter a flight in maximum drag attitude increases the miss distance monotonously up to $ \num{13.30} $ times the predicted miss distance of $ \SI{761.0}{\meter} $. A minimum drag manoeuvre, however, leads to a shrinking miss distance for shorter manoeuvre times. Flying in minimum drag attitude for $ \SI{50.69}{\hour} $ before TCA  minimizes the miss distance to yield only $ \SI{1.964}{\meter} $. This is the result of an achieved separation distance which corresponds to the in-track relative position component in the predicted conjunction geometry (\cf \cref{fig:conjunction_geometries}). For longer manoeuvre duration and therefore higher absolute separation distances the minimum drag manoeuvre is able to increase the miss distance as well. For a minimum drag manoeuvre lasting the whole 5 days a miss distance of $ \SI{2.924}{\kilo\meter} $ can be achieved, which is an increase by $ \SI{284.3}{\percent}$ compared to the original prediction. Looking at scenario B, the situation is different. Here a minimum drag manoeuvre leads to a monotonously increasing miss distance of up to $ \SI{2.868}{\kilo\meter} $ compared to the $ \SI{554.1}{\meter} $ that were originally predicted. On the other hand, a flight in maximum drag attitude for $ \SI{42.64}{\hour} $ minimizes the miss distance to only $ \SI{61.72}{\meter} $, \ie $ \SI{11.14}{\percent} $ of the miss distance for no manoeuvre. However, for longer manoeuvre duration in maximum drag the miss distance increases and finally equals the achievable miss distance via minimum drag for a manoeuvring duration of $ \SI{71.71}{\hour} $. For even longer manoeuvres the miss distance is greater for a flight in maximum drag attitude.

\begin{figure}[h]
	\centering
	\begin{subfigure}{\linewidth}
		\centering
		\resizebox{\linewidth}{!}{\includegraphics{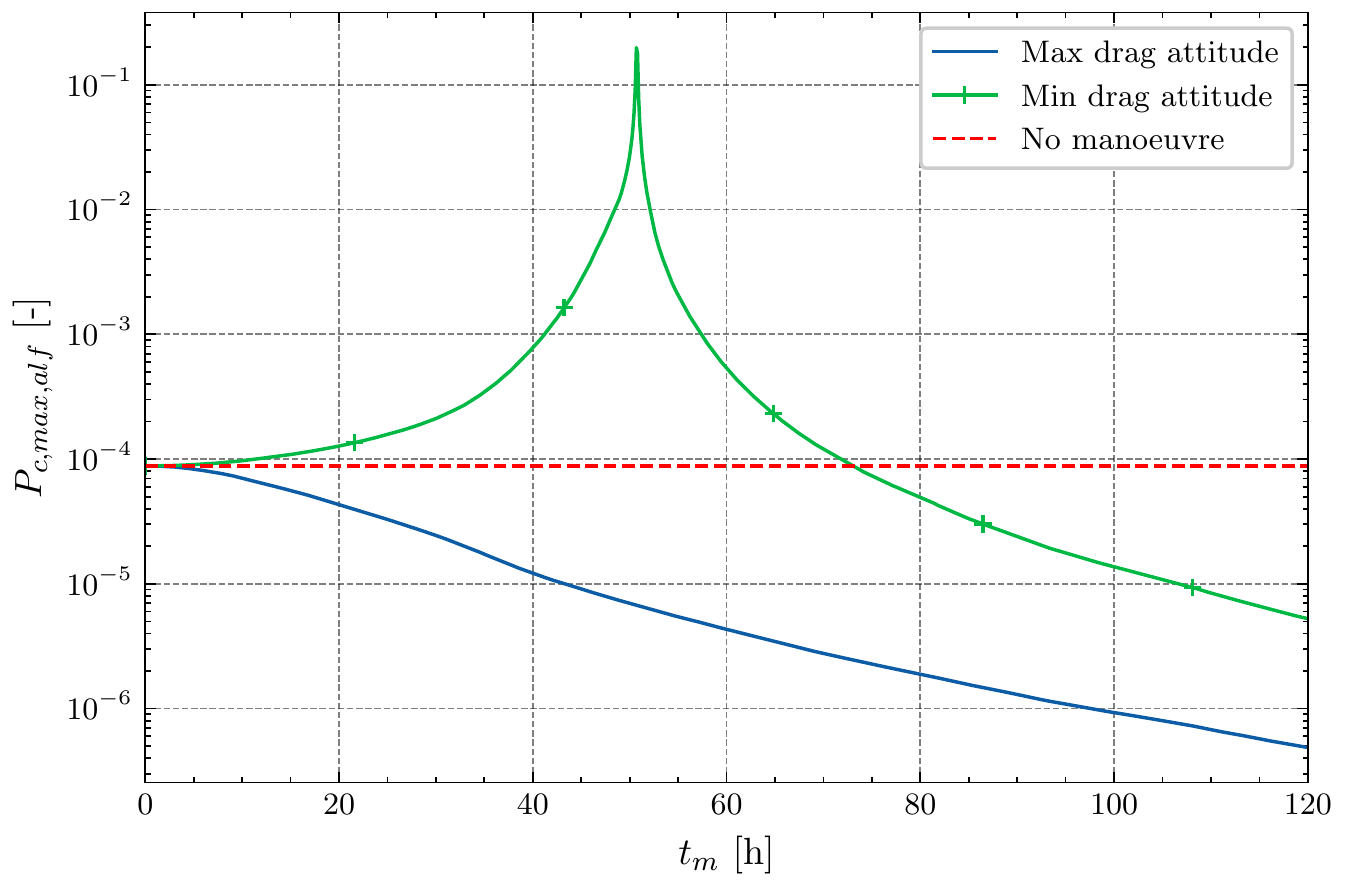}}
		\caption{Encounter A.}
		\label{fig:pc_alfano_a}
	\end{subfigure}\\
	\begin{subfigure}{\linewidth}
		\centering
		\resizebox{\linewidth}{!}{\includegraphics{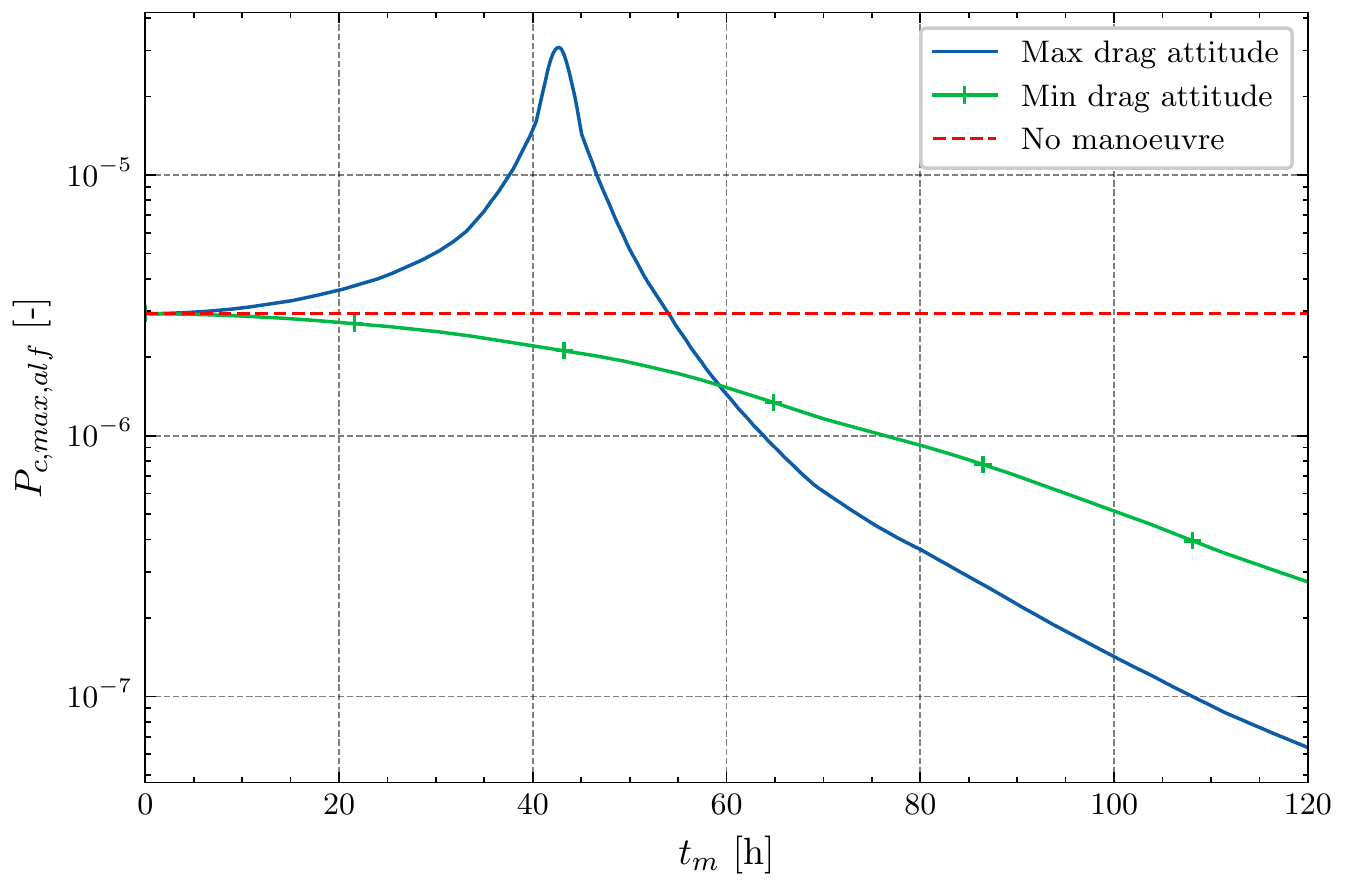}}
		\caption{Encounter B.}
		\label{fig:pc_alfano_b}
	\end{subfigure}
	\caption{Maximum probability of collision depending on manoeuvring time.}
	\label{fig:pc_alfano}
\end{figure}

To compare the risk associated with the manoeuvres, the maximum collision probability is calculated according to Alfano's method (\cf \cref{eq:pc_max_alfano}), as visible in \cref{fig:pc_alfano}. The aforementioned findings settle down in the collision probability as well. The increasing miss distance for a maximum manoeuvre in case A leads to a decreasing collision probability, whereas the minimum drag manoeuvre causes $ P_{c,max,alf} $ to show a maximum at exactly the manoeuvring time minimising the miss distance. For encounter B it is exactly the other way round. A maximum drag manoeuvre shorter than $ \SI{57.50}{\hour} $ leads to a maximum collision probability higher than if no manoeuvre were to be performed, thus worsening the situation. However, the growing achievable miss distances eventually lead to the fact that a manoeuvre in maximum drag attitude causes a lower maximum collision probability than a minimum drag manoeuvre.

\subsection{Influence of parameter uncertainties}
Uncertainties in the parameters of the analytic equation for the achievable separation distance lead to additional in-track position uncertainties at the time of closest approach. This additional uncertainty $ \sigma_{\Delta x} $ can be expressed as a scaling factor for the respective element of the covariance matrix of the manoeuvring satellite at TCA. The updated covariance matrix follows as
\begin{equation}
	\m{C}_{1,tca}' = \mat{1 &0 &0\\
		0 &k^2 &0\\
		0 &0 &1}  \m{C}_{1,tca}
\end{equation}
\begin{equation}
	\text{with\quad} k = \frac{\sigma_{T,1} + \sigma_{\Delta x}}{\sigma_{T,1}}
\end{equation}
where $ \m{C}_{1,tca} $ is the position covariance matrix and $ \sigma_{T,1} $ its in-track component of the satellite as defined in the CDM.
The effect of these uncertainties on the collision probability will be examined based on the exemplary close encounter of the \flp on April 7, 2022. The reasonable CAM is a flight in maximum drag attitude, for which the manoeuvring time will be varied in the following analyses ($ \SI{10}{\hour} $, $ \SI{20}{\hour} $, $ \SI{30}{\hour} $, $ \SI{40}{\hour} $, and $ \SI{50}{\hour} $). For simplicity, no further constraints on the manoeuvre will be considered. At first, the influence of the scaling factor $ k $ on the collision probability will be quantified. Afterwards, it will be examined how uncertainties in the different parameters translate into $ k $ and thus affect $ P_c $.

\begin{figure*}[p!]
	\centering
	\includegraphics{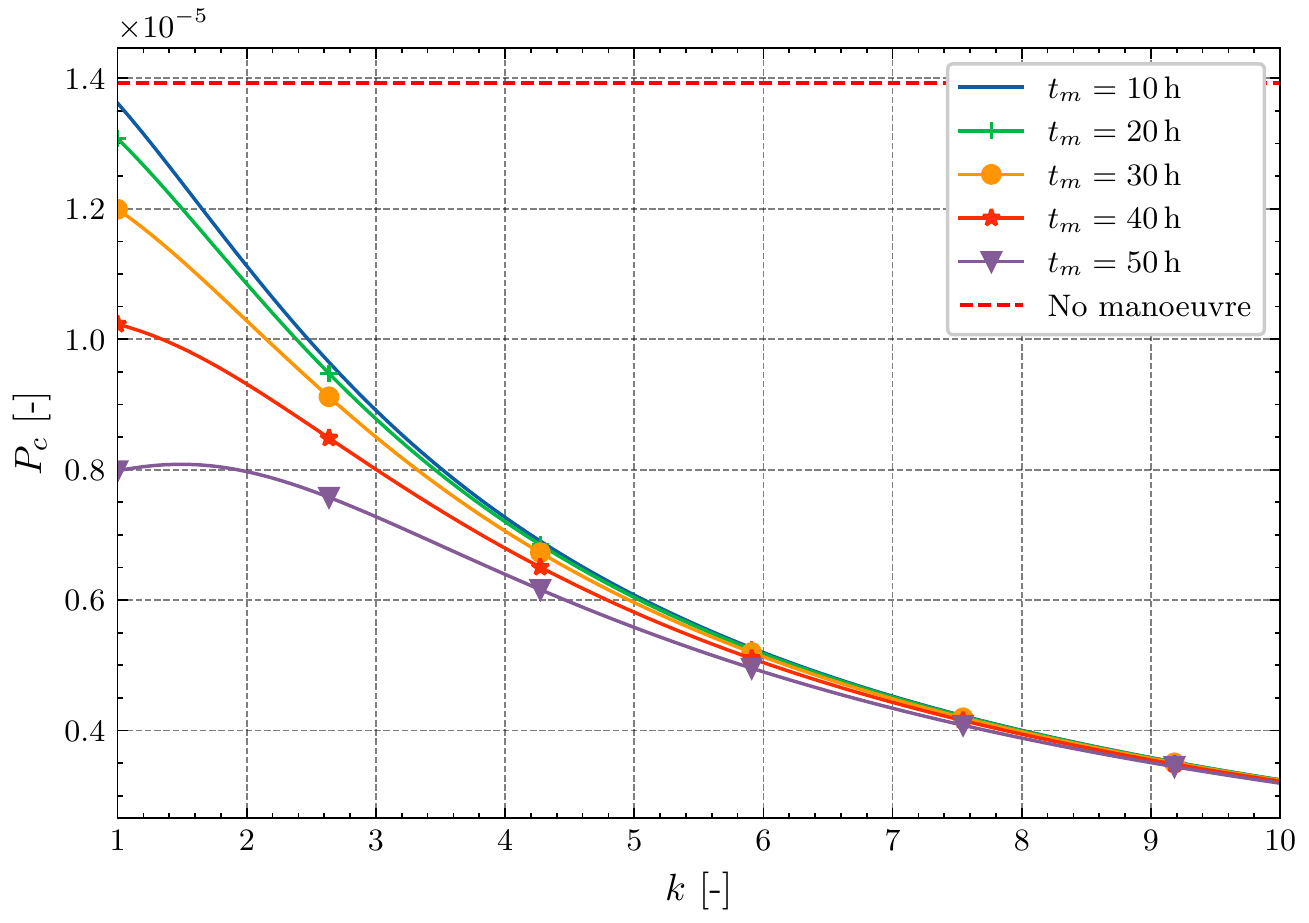}
	\caption{Collision probability over the in-track covariance scaling factor. In red and dashed is the collision probability for no manoeuvre.}
	\label{fig:scaling}
\end{figure*}

In \cref{fig:scaling}, the resulting collision probability is plotted over $ k $ for the different manoeuvring times. If no parameter uncertainties are considered, $ k=1 $ and the collision probability can be decreased compared to the predicted value  for any manoeuvring time $ >\SI{0}{\second} $. The collision probability tends to decrease monotonously with growing scaling factor for the manoeuvring times up to $ \SI{40}{\hour} $. The longest manoeuvre duration of $ \SI{50}{\hour} $, however, shows a maximum collision probability, which is reached for $ k_{max}=1.455 $. This means that parameter uncertainties in the analytic equation, which cause the in-track covariance to grow by $ \SI{45.5}{\percent} $, in fact lead to an increased collision probability compared to when neglecting the parameter uncertainties. It becomes evident as well, though, that the resulting collision probability is still below the one without manoeuvre.

\begin{figure*}[p!]
	\centering
	\includegraphics{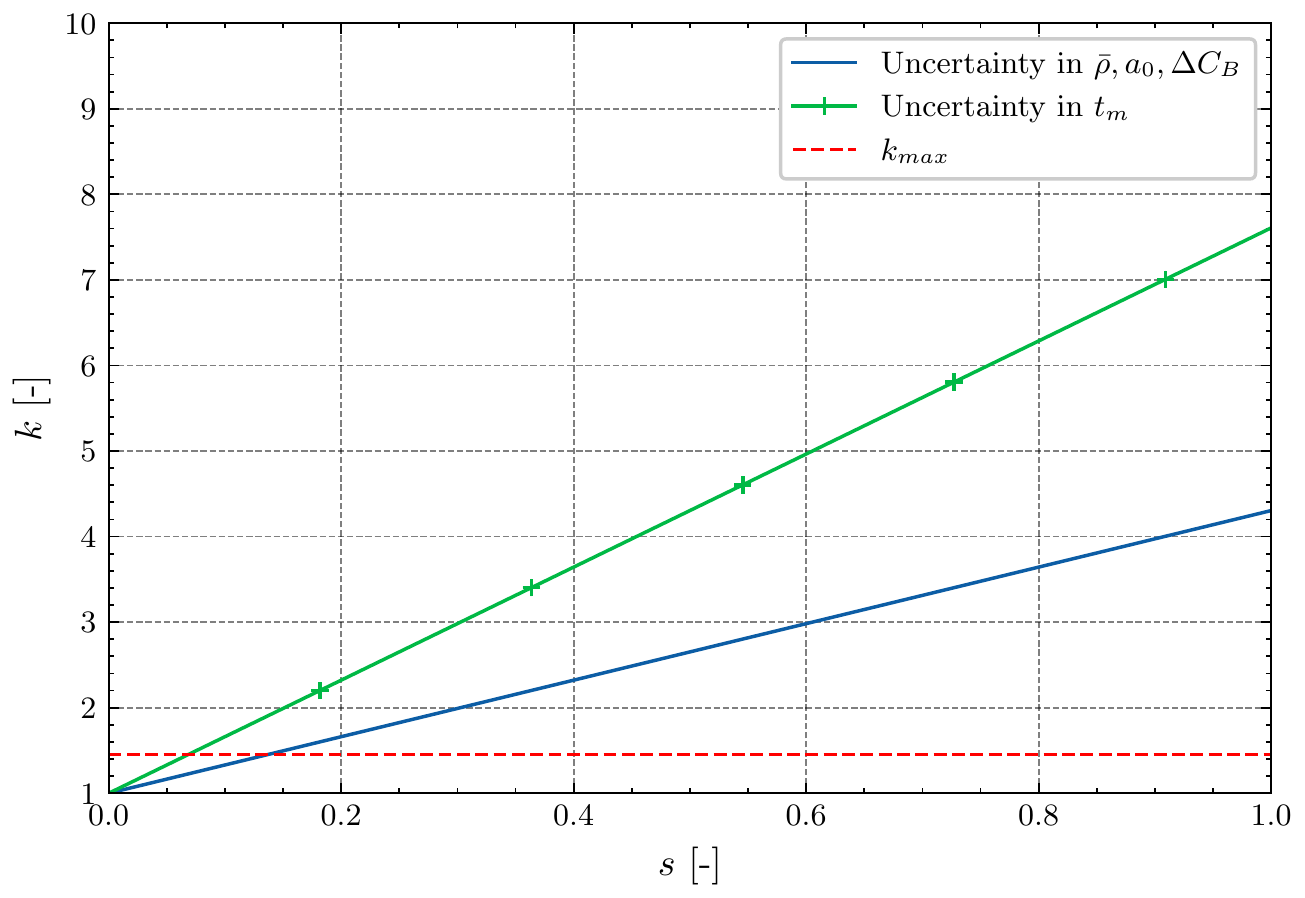}
	\caption{In-track covariance scaling factor depending on an uncertainty level $ s $ in the individual parameters of the analytic equation. The manoeuvring time is $ t_m=\SI{50}{\hour} $. The scaling factor $ k_{max} $, for which $ P_c $ becomes maximum, is marked in red and dashed.}
	\label{fig:k}
\end{figure*}

\cref{fig:k} depicts the scaling factors depending on the uncertainty level $ s $ in the parameters $ \bar \rho $, $ a_0 $, $ \Delta \beta^* $ or $ t_m $ for a manoeuvring time of $ t_m = \SI{50}{\hour} $. Due to the  linear error propagation, the first three parameters show the same effect on the uncertainty in the achievable separation distance and thus on the scaling factor, whereas the impact of an uncertainty in the manoeuvring time is twice as high (\cf \cref{eq:error_propagation_rho,eq:error_propagation_sma,eq:error_propagation_CB,eq:error_propagation_tc}). 
An uncertainty in any one of atmospheric density, semi-major axis, or inverse ballistic coefficient difference of $ \SI{14.14}{\percent} $ results in a scaling factor of $ k_{max} $ leading to the maximum in collision probability. For the uncertainty in the manoeuvring time $ t_m $, this is the case at $ \SI{7.071}{\percent} $.

\section{Discussion}
\label{ch:discussion}
In the following chapter the results of the previous analyses are discussed. At first, the feasibility of aerodynamic collision avoidance manoeuvres for the \flp is evaluated. After that, the effect of uncertainties is dealt with before possible manoeuvre strategies are discussed.

\subsubsection{Feasibility}
\begin{figure}[h]
	\centering
	\resizebox{\linewidth}{!}{\includegraphics{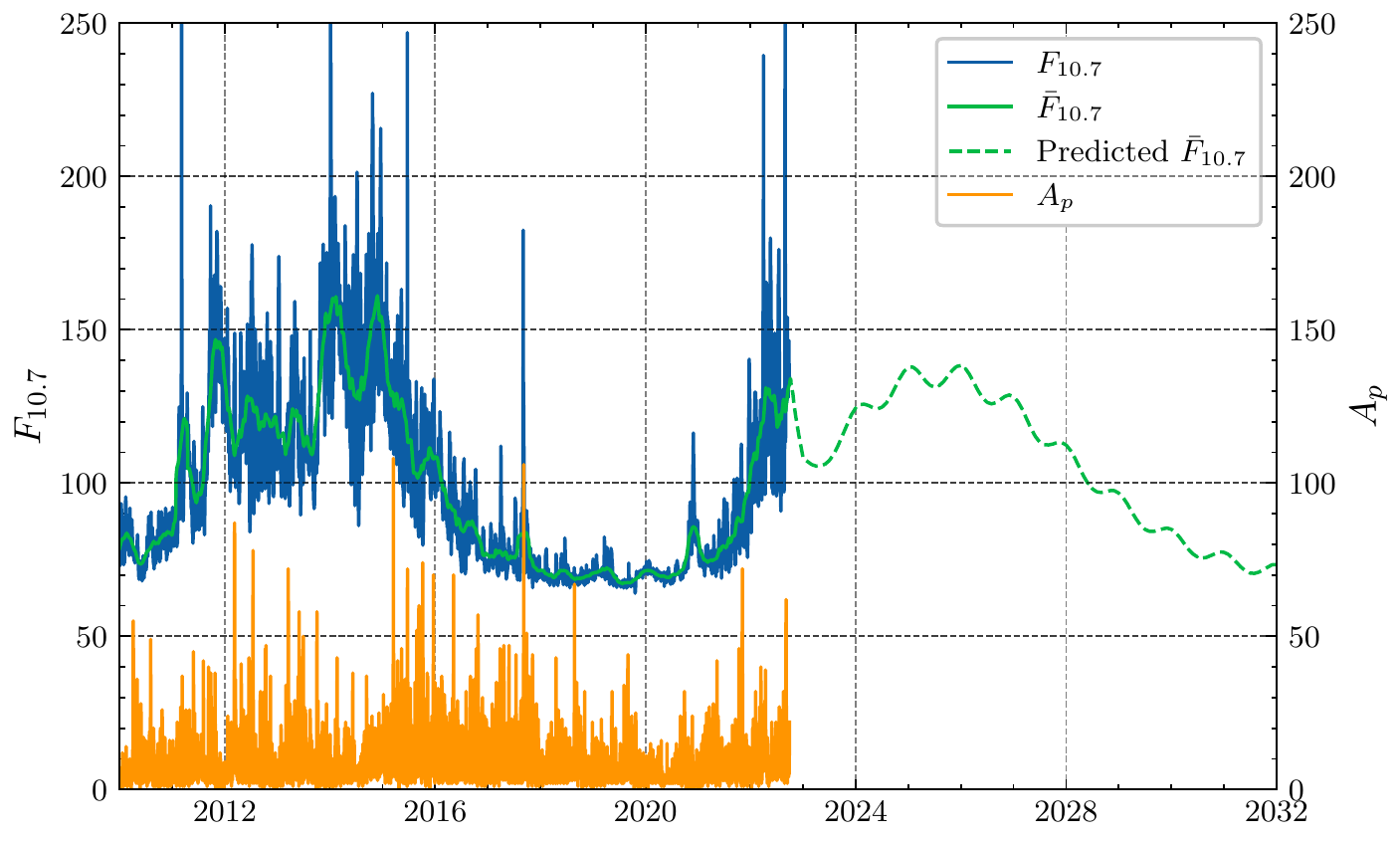}}
	\caption{Solar activity proxy $ F_{10.7} $, its 81-day average $ \bar F_{10.7} $ and geomagnetic activity index $ A_p $ over time. Historic and predicted data are obtained from \citet{SET}.}
	\label{fig:space_weather}
\end{figure}

Atmospheric density is highly dependent on the activity of the Sun and Earth's geomagnetic activity. This already highlights the importance of considering current activity data when estimating effects of aerodynamic manoeuvres. By employing complex atmospheric models, recent and predicted values for the indices and proxies can be used to estimate the density on a given trajectory. In a first step, the \flp's ability to perform CAMs depending on solar and geomagnetic activity was studied. As expected, the achievable separation distance in a given time is strongly influenced by the activity level, leading to separation distances which vary across two orders of magnitude.
\citet{Mishne.2017} argued that for a reasonable collision avoidance manoeuvre, the achievable separation distance must be in the range of $ \SI{900}{\meter} $ per three days in order for it to be greater than typical propagation uncertainties and thus useful. It can be concluded that the \flp can achieve this separation distance for moderate and high levels of solar and geomagnetic activity. Only for low activity, the resulting forces due to aerodynamic drag appear to be too low for creating significant separations. Solar activity follows approximately an 11-year cycle and the next maximum is expected for 2025 \cite{Hathaway.2015, NOAA.2019}. \cref{fig:space_weather} presents the course of the solar flux at $ \SI{10.7}{\centi\meter} $ as proxy for the Sun's activity. This allows to conclude that the feasibility of aerodynamic CAMs will be given in the upcoming period until the next minimum in the solar cycle. During minimum phases, CAMs using aerodynamic drag might temporarily not be feasible for the \flp, as long as solar activity and consequently atmospheric density is low.
It is to note, that these results strongly depend on the ballistic coefficient. The used ballistic coefficients are considered lower limits. hence, the achievable separation distances are lower limits as well and the absolute values might be lower for a minimum drag maneouvre. Here, further analysis of the error in the determined ballistic coefficient is necessary.

Considering constraints for constraint phases, \ie charging phases here, shows an influence on the achievable separation distance. It was observed, though, that even for a manoeuvre in which the \flp holds maximum drag attitude for $ \SI{1.5}{\hour} $ followed by a charging phase of $ \SI{2.5}{\hour} $ the resulting separation distance fulfills the criterion mentioned before and can therefore be considered a useful manoeuvre. Considering very short manoeuvres under $ \SI{24}{\hour} $, a split of $ \SI{2}{\hour} $ manoeuvring and $ \SI{2}{\hour} $ charging is necessary for an achieved separation distance of $ \SI{300}{\meter} $ within the first day. Such constraints, however, are far beyond what is typically required for charging phases.

In the context of the \flp's orbital evolution it is to note that the angle between the orbital plane and the Sun vector is continuously changing. The recent state demands charging phases as mentioned before when performing maximum drag manoeuvres. Minimum drag manoeuvres can be implemented without having to consider charging phases, since the satellite can be freely moved around the axis pointing to the direction of flight, such that the solar arrays can point to the Sun constantly. For a progressively decreasing local time of ascending node, this situation will change. For a LTAN around midnight (or noon, respectively), the orbital plane is oriented in a way allowing to point the solar arrays to the Sun during a maximum drag manoeuvre. In this case, charging phases were to be implemented during flights in minimum drag attitude. Looking at orbital plane orientations between these extreme cases, further tests will be necessary to determine whether enough power can be generated in the respective attitudes or whether charging phases are needed. Furthermore, tests should be performed in order to study the effects of uneven heating of the satellite structure, which may, in fact, deteriorate battery performance as well or lead to other negative effects. Still, this can be mitigated by making use of the remaining degree of freedom and commanding rotations around the direction of flight without worsening the manoeuvring performance.

Power could be saved by minimizing downlinks of the satellite, \eg by not establishing connection to the ground station during a pass at all or by downlinking fewer data.
However, monitoring the satellite's telemetry data is an important point when performing collision avoidance manoeuvres. Problems with the GPS receivers or the star trackers, for example, could potentially take the satellite its ability to attain commanded attitudes. Since this might worsen the situation at TCA, it is recommended to constantly monitor both telemetry as well as housekeeping data, like battery voltages, to ensure that the satellite performs the manoeuvre as desired.

\subsection{Manoeuvre suggestions}
When facing a predicted close encounter, the operators of the \flp so far have three options. They can either command a flight in maximum or in minimum drag attitude or do not implement any CAM at all. Depending on the predicted collision probability, performing a CAM and thus reducing the risk is desirable, however. In \cref{sec:ana_conjunction_geometry} it was found that a positive separation distance (corresponding to a flight in maximum drag attitude) is desirable if the predicted in-track component of the relative position at TCA is negative. Vice versa, a negative separation distance is to be induced for a positive predicted in-track distance at TCA. For longer manoeuvre durations, though, it may be favourable to perform a contrary manoeuvre. Depending on the reference ballistic coefficient (as well as solar and geomagnetic activity) it might be possible to implement a manoeuvre which would lead to a shrinking miss distance if performed shortly but to drastically growing miss distances for longer manoeuvring times, as was the case in encounter B in \cref{sec:ana_conjunction_geometry}.
From this, possible manoeuvring strategies might be deduced. For short-term manoeuvres with little time left until predicted TCA the attitude which monotonously increases the miss distance is to be commanded, representing a conservative. This results in a maximum drag attitude if the in-track component of the relative position of the secondary object with respect to the manoeuvring satellite is negative and a minimum drag attitude if it is positive. For long-term manoeuvres, the strategy might be inverted if an analysis of the achievable separation and miss distances indicates so. At this point in time it can, however, not be recommended to perform a manoeuvre which leads to decreasing miss distances at first. Further analyses on the accuracy of the predicted separation distances and evaluation of actually performed CAMs are necessary before such a manoeuvre might be reliably performed.

In general, it can be concluded that an early implementation of a manoeuvre is more effective. In \cref{sec:manoeuvre}, it was shown that the early build-up of angular motion relative to the reference trajectory leads to an increasing separation distance even though the satellite might return to its reference attitude, which would be the case if the CAM were, for example, ended and the satellite returned to nominal operation. From another point of view, an early manoeuvre implementation leads to a shorter overall time spent in the manoeuvring attitude to achieve a specific separation distance compared to if the manoeuvre was to be performed at a later point.
For charging phases, this is already considered in the analysis tool, since they follow manoeuvring phases.

Another point was not addressed so far. For a secondary object in an orbit with a similar period to that of the satellite, it is realistically possible for further encounters to happen several orbits before and after the predicted close encounters. It was so far not analysed, how avoidance manoeuvres using aerodynamic drag affect previous or later encounters. This is an important aspect of future research. Whether sgp4-propagation of TLEs is sufficient for a reliable simulation is difficult to assess. High-fidelity orbit propagation might lead to further insights.

\subsection{Uncertainties in parameters}
The effects of uncertainties in the parameters required for the equation to estimate the achievable separation distance were investigated on an exemplary close encounter. Uncertainties in the average density, the semi-major axis and the change in inverse ballistic coefficient were covered. Using a linear error progression, it was calculated to which extent they affect the uncertainty in the separation distance. This uncertainty adds to the in-track component of the position covariance as given in CDMs.
For the exemplary encounter it was found that, depending on the manoeuvring time, uncertainties in the parameters affect the resulting collision probability differently. For shorter manoeuvres the collision probability strictly decreases for growing parameter uncertainties. On the other hand, the collision probability shows a maximum for a certain scaling factor of the in-track position covariance before strictly decreasing as well. The scaling factor to reach maximum collision probability translates into uncertainties of about $ \SI{10}{\percent} $ for either averaged density, semi-major axis, or change in inverse ballistic coefficient. Alternatively, a manoeuvring time with an uncertainty of about $ \SI{5}{\percent} $ has the same effect. Taking into account, that these parameters are all fraught with uncertainties and that these uncertainties add up, the individual uncertainties must even be lower to maximise $ P_c $. Still, it becomes clear that, in the case at hand, for any additional uncertainty the collision probability is clearly lower than the one which was previously calculated in the case of not manoeuvring due to the positive impact of the increased miss distance.

\subsection{Uncertainty quantification}
Looking at the individual parameters, the uncertainty levels may be assessed. All the findings are summarized in \cref{tab:uncertainties}.
Atmospheric density varies over two orders of magnitude at the altitude regime of the \flp, when comparing low to high solar and geomagnetic activity. The uncertainty in density is, thus, a major factor and caused by different aspects. First of all, the atmospheric model in use provides only estimations of the density. While no model provides overall better estimations than another, an average uncertainty of $ \SIrange{10}{15}{\percent} $ needs to be assumed according to literature \cite{Sagnieres.2017, Vallado.2014b}. For short-term predictions uncertainties might even be significantly higher, which leads to the next aspect. The implemented atmospheric models need indices and proxies for solar and geomagnetic activity as input, which have to be predicted in order to estimate atmospheric densities at future points in time. \citet{Vallado.2014b} analysed solar flux predictions and found that they can show significant error when compared to the actual observations, especially during solar maxima. Conclusively, they stated a standard deviation of $ \SIrange{20}{40}{} $ solar flux units for 45-day forecasts, leading to density variations of $ \SIrange{150}{300}{\percent} $. The geomagnetic activity  was found to be accurate within $ \SI{20}{} $ units, resulting in a density error of up to $ \SI{50}{\percent} $. Regarding the error in atmospheric density induced through positional errors due to sgp4-propagation, a first estimation has been established. A \flp trajectory was propagated for five days using sgp4 and a high-fidelity orbit propagation and the encountered densities were each averaged. The difference was found to be $ <\SI{1}{\percent} $ \cite{Turco.2022}.

The semi-major axis is obtained from TLEs. Looking at recent TLEs for the \flp, the obtained semi-major axis showed variations of up to $ \SI{50}{\meter} $.
Compared to the JSpOC observations, though, differences of up to $ \SI{10}{\kilo\meter} $, \ie $ \SI{0.1435}{\percent} $, were found.

The accuracy of the change in inverse ballistic coefficient is dependent on the accuracy of the reference coefficient in the CDM and the accuracy of the inverse ballistic coefficient in manoeuvring attitude. Currently, no information on potential errors of the reference inverse ballistic coefficient is available. Regarding the manoeuvring attitude, two aspects influence the accuracy of the ballistic coefficient. First, the aerodynamic analysis of the \flp, especially with the assumptions on gas-surface interaction introduces error. Its magnitude can so far not be assessed but the determined inverse ballistic coefficients are expected to be lower bounds, leading to the resulting separation distances being lower bounds with regards to $ \beta^* $ as well. The second aspect is the pointing accuracy of the \flp. Deviations from the assumed attitude affect the surface exposed to the flow, the angle of incidence of the atmospheric particles and thus the inverse ballistic coefficient. The pointing error is stated to be within $ \SI{0.042}{\degree} $ \cite{Eickhoff.2016}, which translates to an error in the inverse ballistic coefficient of $ \le \SI{5}{\percent} $.

Uncertainties in the manoeuvring time are not quantified at the moment. With a rotation speed of $ \SI{1.5}{\degree\per\second} $, the \flp is able to turn into any specified attitude within $ \SI{120}{\second} $ and this can be considered in the manoeuvring time itself. Losses in manoeuvring time can further happen due to the \flp returning to a safe mode losing star tracker functionality or similar events, which cause an attitude potentially different from the commanded one. To represent such events statistically, further research needs to be conducted on their frequency.

The results indicate the importance of estimating uncertainties in the parameters as realistically as possible to generate a trustworthy probability of collision. Under or over-estimating the uncertainties may lead to increased or decreased collision probabilities. This does not only include the parameter uncertainties but also the position covariance given in CDMs and further scaling by JSpOC, which is another source for a variation of the collision probability. Furthermore, correlations between the CDM covariance and the uncertainties in the parameters, which are so far assumed to be zero, should be investigated more deeply to obtain a realistic position covariance after a performed manoeuvre. This directly increases the fidelity of the determined collision probability. So far, the most reliable possibility for assessing the overall risk of a close encounter is the calculation of a maximum collision probability, which is detached from the objects' covariances in its worst-case formulation. While the collision probability itself might be mismodelled, the maximum collision probability shows a positive influence of any increase in miss distance through an aerodynamic manoeuvre, which only depends on the achievable separation distance. 

\begin{table*}[h!]
	\centering
	\caption{Causes for uncertainties in the parameters of the analytic equation for the achievable separation distance.}
	\label{tab:uncertainties}
	\begin{tabular}{@{}llcc@{}}
		\toprule
		Parameter &Cause of uncertainty &Expected uncertainty &\begin{tabular}[c]{@{}l@{}}Expected uncertainty\\in the parameter\end{tabular} \\
		\midrule
		\multirow{4}{*}{Density $ \bar \rho $} &Atmospheric model accuracy &$ \SIrange{10}{15}{\percent} $ \cite{Vallado.2014b} & $ \SIrange{10}{15}{\percent} $\\
		&Solar flux activity prediction &$\le \SIrange{20}{40}{units} $ \cite{Vallado.2014b} & $ \le \SIrange{150}{300}{\percent} $\\
		&Geomagnetic activity prediction &$ \le \SI{20}{units} $ \cite{Vallado.2014b} &$ \le \SI{50}{\percent} $\\
		&sgp4-propagation and density averaging &$ \le\SI{1}{\percent} $\textsuperscript{1} &$ \le\SI{1}{\percent} $\\
		\midrule[0.25pt]
		Semi-major axis $ a_0 $ &TLE accuracy &$ \le \SI{10}{\kilo\meter} $\textsuperscript{1} &$ \le \SI{1}{\percent} $\\
		\midrule[0.25pt]
		\multirow{3}{*}{\begin{tabular}[c]{@{}l@{}}Change in inverse ballistic\\coefficient $ \Delta \beta^* $\end{tabular}} &Reference inverse ballistic coefficient $ \beta^*_{ref} $ &? &?\\
		&Aerodynamic analysis &? &?\\
		&Pointing accuracy &$ \SI{0.042}{\degree} $ \cite{Eickhoff.2016} &FLP: $ \le \SI{5}{\percent} $ \\
		\midrule[0.25pt]
		Manoeuvring time $ t_m $ &Safe mode, loss of star trackers, etc. & - & -\\
		\bottomrule
	\end{tabular}
	\textsuperscript{1} This is an estimate based on first analyses with the \flp and requires further research.
\end{table*}
\section{Summary}
\label{sec:summary}
In this work, a tool was developed to analyse aerodynamic manoeuvres for satellite collision avoidance. Different manoeuvres can be compared with respect to their resulting miss distance and collision probability, considering solar and geomagnetic activity conditions. 
Further constraints regarding charging phases during the manoeuvre can be considered and uncertainties in the various parameters that are used as input to the tool can be accounted for, as well. At this point, the tool relies solely on CDM data and (publicly) available TLEs, making it simple to use. No further orbit determination and propagation processes are required to estimate the effect of a manoeuvre.

Close encounters of the university satellite \flp were used as exemplary cases. The use of aerodynamic manoeuvres for collision avoidance was found to be feasible for the \flp for moderate and high solar and geomagnetic activity levels. Implementing charging phases of reasonable duration were shown to not hinder the applicability of aforementioned manoeuvres. First manoeuvre suggestions based on the conjunction geometry were deducted. Additionally, an exemplary conjunction was studied to evaluate the influence of parameter uncertainties. For this example, it was concluded that while the uncertainties may lead to slightly higher collision probabilities than when not accounting for them, the overall collision probability is decreased with a manoeuvre.

\section{Outlook}
\label{sec:outlook}
More research is necessary for a realistic approximation of the uncertainties in the parameters, their correlation to the position uncertainty defined in a CDM to obtain a reliable and less conservative collision probability. The uncertainty sources need to be addressed individually and their effect on the manoeuvre uncertainty studied.

At the Institute of Space Systems, University of Stuttgart, the \flp further offers the possibility to verify the effect of aerodynamic manoeuvres. In future work, flight tests can be performed and analysed. Upcoming other missions can benefit from insights and the use of the analysis tool as well.


\appendix

\section{Space weather parameters}
For analyses, levels of solar and geomagnetic activity are used as defined in ISO1422 \cite{ISO1422}. \cref{tab:activity_levels} shows the respective values.
\begin{table}[H]
	\centering
	\caption{Levels of solar and geomagnetic activity as defined in ISO1422 \cite{ISO1422}. Each value is given in the respective index unit and is to be applied for the average index value as well. A further high short-term level is given in the standard, which is not included here.}
	\label{tab:activity_levels}
	\begin{tabular}{@{}lcccccc@{}}
		\toprule
		Activity level &$F_{10.7}$ &$S_{10}$ &$M_{10}$ &$Y_{10}$ &$A_p$ &$Dst$\\
		\midrule
		Low &65 &60 &60 &60 &0 &-15\\
		Moderate &140 &125 &125 &125 &15 &-15\\
		High &250 &220 &220 &220 &45 &-100\\
		\bottomrule
	\end{tabular}
\end{table}

\section{Aerodynamic analysis of the \flp}
\label{sec:aerodynamic_analysis}

The ballistic coefficient of a simplified model of the \flp was determined for different attitudes using the publicly available software tool ADBSat developed by \citet{ADBSatMethodology.2022}. 

\begin{figure}[h]
	\centering
	\begin{subfigure}[c]{0.3\linewidth}
		\centering
		\resizebox{\linewidth}{!}{
		\includegraphics{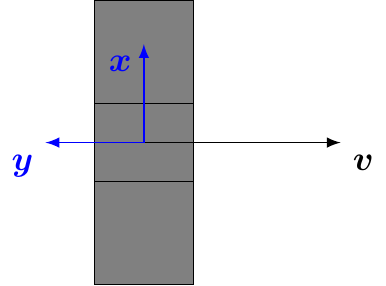}}
		\caption{Minimum drag attitude.}
		\label{fig:minimum_drag_attitude}
	\end{subfigure}\hfill
	\begin{subfigure}[c]{0.3\linewidth}
		\centering
		\resizebox{\linewidth}{!}{
%
		\includegraphics{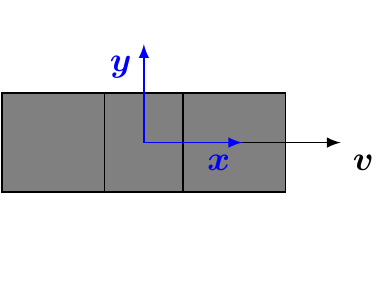}}
		\caption{Nadir-pointing attitude.}
		\label{fig:nadir_pointing_attitude}
	\end{subfigure}\hfill
	\begin{subfigure}[c]{0.3\linewidth}
		\centering
		\resizebox{\linewidth}{!}{
		\includegraphics{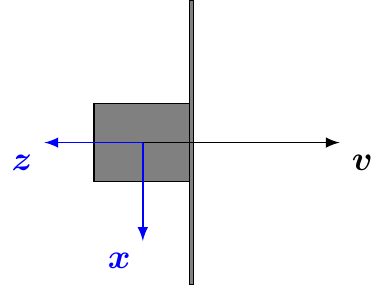}}
		\caption{Maximum drag attitude.}
		\label{fig:maximum_drag_attitude}
	\end{subfigure}
	\caption{Schematic illustrations of the different specified attitudes of the \flp with the body coordinate system. All attitudes can be rotated around the direction of flight without changing the flow field.}
	\label{fig:attitudes}
\end{figure}

The ballistic coefficient of the \flp can be maximized by maximizing the area perpendicular to the flight velocity, corresponding to a flight with maximum drag. Vice versa, the minimization of this area leads to a minimum ballistic coefficient. The respective maximum and minimum drag attitudes are defined by having either the body $ z $-axis (maximum drag) or $ y $-axis (minimum drag) point parallel to the direction of flight. To protect the payload cameras and the star trackers from atmospheric particles, the negative $ z $ or $ y $-axes are pointed towards flight direction. The Nadir-pointing attitude is defined by the $ z $-axis pointing towards the centre of Earth and the $ x $-axis in the direction of flight. The different attitudes are shown in \cref{fig:attitudes}.

An approximated circular orbit of the \flp is assumed with an inclination of $ \SI{97.4}{\degree} $ and a local time of ascending node of 06:00 h. The ADBSat tool is employed to determine the ballistic coefficient of the satellite in the respective attitudes at 100 points on this orbit and the results are then averaged over these evaluation points.
Further, levels of solar and geomagnetic activities, defined by the solar flux, its 81-day average and the planetary equivalent amplitude, are varied. The resulting ballistic coefficients are stored in 3-dimensional look-up tables, allowing a determination of the ballistic coefficient of an attitude via linear interpolation given specific values of the activity.

\cref{tab:aero_results} provides the resulting ballistic coefficients for different activity levels (\cf \cref{tab:activity_levels}). 

\begin{table}[htb!]
	\centering
	\caption{Inverse ballistic coefficient $ \beta^* $ [\SI{}{\square\meter\per\kilogram}] of the \flp for different attitudes and different levels of solar and geomagnetic activity.}
	\label{tab:aero_results}
	\begin{tabular}{@{}lccc@{}}
		\toprule
		&\multicolumn{3}{c}{$ \beta^* $ [\SI{}{\square\meter\per\kilogram}]}\\
		\midrule
		Activity level&Low&Moderate&High\\
		\midrule
		Minimum drag &$ \SI{1.384e-2}{} $ &$ \SI{1.214e-2}{} $ &$ \SI{1.220e-2}{} $ \\
		Nadir-pointing &$ \SI{1.475e-2}{} $ &$ \SI{1.324e-2}{} $ &$ \SI{1.328e-2}{} $ \\
		Maximum drag &$ \SI{3.377e-2}{} $ &$ \SI{3.262e-2}{} $ &$ \SI{3.258e-2}{} $ \\
		\bottomrule
	\end{tabular}
\end{table}

Gas-surface interactions are modelled using the Sentman model \cite{ADBSatMethodology.2022,Moe.2005,Sentman.1961}, which assumes diffuse re-emission and is usually used for satellites at lower orbital altitudes than the \flp. With increasing altitude, particle interaction with the surface tends to shift to specular reflection and thus higher drag coefficients. Hence, the results presented here are considered lower limits.

\bibliographystyle{elsarticle-num-names} 
\bibliography{literature}





\end{document}